\documentclass[table]{article} 
\usepackage{main}
\usepackage{soul} 
\usepackage{booktabs}
\usepackage{graphicx}
\usepackage{enumitem}
\usepackage{wrapfig}
\usepackage{algorithm}
\usepackage{algpseudocode}
\usepackage{adjustbox}
\usepackage{hyperref}
\usepackage{microtype}
\usepackage{multicol}
\usepackage{float}
\usepackage{amsmath}
\usepackage{amssymb}
\usepackage{colortbl}
\usepackage[utf8]{inputenc}
\definecolor{lightgray}{rgb}{0.9,0.9,0.9}
\usepackage{caption}
\usepackage{subcaption}
\usepackage{xcolor}
\usepackage{setspace}
\usepackage{url}
\usepackage{multirow}
\usepackage{colortbl}
\usepackage{tabularx}
\usepackage{blindtext}
\usepackage{pgfplots}
\pgfplotsset{compat=1.18} 
\usepackage{tikz}
\usetikzlibrary{er,positioning,bayesnet}
\usepackage{makecell}
\usepackage{tipa}
\usepackage{siunitx}
\usepackage{nicefrac}
\usepackage{tocloft}
\usepackage{listings}
\usepackage[raster,skins]{tcolorbox} %
\usepackage{xltabular}
\usepackage{adjustbox}
\usepackage{xurl}
\usepackage{CJKutf8} 
\usepackage{tcolorbox}
\tcbuselibrary{breakable}
\tcbset{
    promptbox/.style={
        colback=cyan!5!white,
        colframe=cyan!50!black,
        fonttitle=\bfseries\large,
        breakable
    }
}

\usepackage{amsmath,amsfonts,bm}

\def\eqref#1{equation~\ref{#1}}
\def\1{\bm{1}}

\DeclareMathAlphabet{\mathsfit}{\encodingdefault}{\sfdefault}{m}{sl}
\SetMathAlphabet{\mathsfit}{bold}{\encodingdefault}{\sfdefault}{bx}{n}

\makeatletter
\newcommand{\affiliation}[1]{\gdef\@affiliation{#1}}
\def\@maketitle{\vbox{\hsize\textwidth
  {\centering{\Large\bf \@title\par}}
  \vspace{0.3cm}
  {\centering\@author\par}     % 去掉 quote 环境
  \vspace{0.1cm}               % 控制 author 和 affiliation 之间的间距
  {\centering\@affiliation\par}% 去掉 quote 环境
  \vspace{0.5cm}
}%
\thispagestyle{firstpage}}
\makeatother

\newcommand{\titlelogo}{%
  \adjustbox{valign=c}{\includegraphics[height=1.4em]{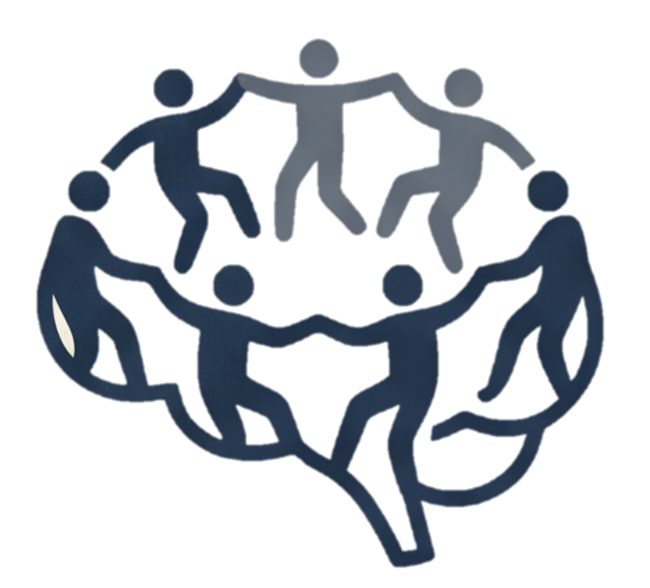}}%
}
\title{%
  \texorpdfstring{\protect\titlelogo\,}{}%
  AgentDisCo: Towards Disentanglement and Collaboration in Open-ended Deep Research Agents%
}

\author{\textbf{
Jiarui Jin, Zexuan Yan, Shijian Wang, Wenxiang Jiao, Yuan Lu
}}

\affiliation{\textbf{
Xiaohongshu Inc.
}}

\usepackage{fontawesome5} % 提供 \faGithub 图标；不想装可去掉并删除下面的 \faGithub
\newcommand{\githuburl}{https://agentdisco-project.github.io/}   % TODO: 改成你的链接
\newcommand{\blfootnote}[1]{%
  \begingroup
    \renewcommand\thefootnote{}\footnote{#1}%
    \addtocounter{footnote}{-1}%
  \endgroup
}
\newcommand{\githubfootnote}{%
  \blfootnote{%
    Project page:
    \href{\githuburl}{\texttt{\githuburl}}%
  }%
}
\newcommand{\correspondencefootnote}{%
  \blfootnote{%
    Correspondence: \texttt{jinjiarui@xiaohongshu.com}, \texttt{wenxiangjiaonju@gmail.com}, \texttt{luyuan3@xiaohongshu.com} 
  }%
}

\begin{document}

\maketitle
\githubfootnote
\correspondencefootnote

\vspace{10pt}

\begin{abstract}
\textbf{Open-ended deep research} agents have emerged as a promising paradigm for autonomously performing comprehensive information gathering and synthesis.
However, existing approaches typically integrate information \textbf{exploration} and \textbf{exploitation} into a single unified module—such as an outline generator or a report generator—thereby limiting their flexibility and optimization potential.

In this paper, we introduce \textbf{AgentDisCo}, a novel \textbf{Dis}entangled and \textbf{Co}llaborative agentic architecture that formulates deep research as an adversarial optimization problem between information exploration and exploitation. 
Specifically, a \textbf{critic agent} is optimized to evaluate and critique the generated outlines (serving as information exploitation states) and subsequently \textbf{refine the search queries (serving as information exploration states)}, while a \textbf{generator agent} is optimized to retrieve updated search results based on the refined search queries (serving as information exploration states) and accordingly \textbf{update the generated outlines (serving as information exploitation states)}.
The resulting outline, progressively refined through iterative adversarial optimization, is subsequently delivered to a downstream report writer module. 
This module leverages the structured outline alongside the accumulated search results to synthesize a comprehensive, coherent, and well-grounded research report.

The above agentic workflow can be optimized through either handcrafted or automatically discovered design strategies by constructing a \textbf{meta-optimization harness} over the adversarial optimization loop, where the generator agent originally tasked with producing target outlines is repurposed as a scoring agent that evaluates and generates quality signals over the critic agent's outputs, thereby enabling systematic optimization of the search queries. 
Concretely, powerful code-generation agents—such as Claude-Code or Codex—are employed to systematically explore the space of agent configurations and automatically construct a \textbf{policy bank}, a structured repository of reusable and composable design strategies over search query generation across diverse research tasks and search domains, enabling the framework to self-refine its own design strategies without requiring extensive human intervention. 
We evaluate AgentDisCo on three widely adopted deep research benchmarks—DeepResearchBench, DeepConsult, and DeepResearchGym—with Gemini-2.5-Pro as our base model, demonstrating performance comparable to or surpassing that of leading closed-source deep research agents.

Furthermore, we observe that existing benchmarks predominantly focus on academic or domain-specific consulting queries, which diverge significantly from the breadth and diversity of real-world user needs. 
To bridge this gap, we introduce \textbf{GALA} (\textbf{G}eneral \textbf{A}I \textbf{L}ife \textbf{A}ssistants), a novel benchmark constructed via an agentic workflow that automatically mines latent deep research interests from users' historical browsing behavior, enabling a more faithful reflection of organic, everyday information needs.

As an intuitive and user-friendly interface is essential for bridging the gap between research outputs and end-user consumption, we develop a \textbf{rendering agent} capable of transforming structured research reports into visually rich (rednote-style) poster presentations. 
Building upon this, we further construct a product demonstration—``AutoResearch Your Interest''—which automatically curates and delivers personalized deep research recommendations tailored to individual user profiles derived from their browsing histories. 
We publicly release our benchmark, code, demo, and evaluation harness to support and accelerate future research in open-ended deep research.
\end{abstract}

% \vfill
% \vspace{2cm}

% \begin{figure}[hbp]
%     \centering
%     \includegraphics[width=0.92\textwidth]{figures/performance.pdf}
%     \caption{Performance and cost comparison of open MoE and dense language models. Circles ($\circ$) denote dense models, while diamonds ($\diamond$) denote MoE models. We benchmark model capabilities using MMLU-Pro, showing that \redmoe achieves comparable accuracy to leading models.}
%     \label{fig:intro}
% \end{figure}

% \vfill

\newpage
\section{Introduction}
\label{sec:intro}
Open-ended deep research agents—capable of synthesizing vast web-scale information into comprehensive, well-cited reports—have emerged as a critical frontier for large language models (LLMs). 
On one hand, closed-source commercial offerings—such as GPT Deep Research \citep{dr} and Gemini Deep Research \citep{GeminiResearch}—provide neither technical reports nor open implementations.
On the other hand, a growing body of open technical reports \citep{han2025deepresearchertesttimediffusion,li2025webweaverstructuringwebscaleevidence,lei2025rhinoinsight} shed light on architectural design choices.
Yet, beneath these advances in released architectural designs lies a persistent architectural bottleneck.
First, existing deep research agents entangle \textbf{information exploitation}—the generation of structured outlines or reports—with \textbf{information exploration}—the planning and generation of search queries—into a single, undifferentiated module, fundamentally lacking any guarantee of informational incrementality in the iterative optimization process.
Second, current outline-guided iteration loops require LLMs to refine generated outlines without explicit optimization objectives, leaving the model without clear guidance on which parts of the outline are satisfactory and which require further improvement. 
This absence of structured feedback signals results in directionless and unstable iterative refinement, where the model oscillates between over-revision and under-revision without convergence.

In this paper, we propose \textbf{AgentDisCo}, a novel \textbf{Dis}entangled and \textbf{Co}llaborative agentic architecture that formulates deep research as an adversarial optimization problem between information exploration and exploitation. 
We first argue that iterative optimization should operate on intermediate generated outlines rather than final reports, as outline-level representations offer greater structural flexibility and are more amenable to effective context management. 
Within the outline optimization loop, we further disentangle the generation of outlines and search queries into two specialized yet interacting agents: a \textbf{critic agent} and a \textbf{generator agent}. 
Specifically, the critic agent receives the current \textbf{generator state (i.e., the information exploitation state)} — comprising the evolving outlines along with their associated references — and is tasked with evaluating and critiquing the quality and completeness of the generated outlines, upon which it subsequently produces targeted and gap-aware search queries.
To further structure this iterative optimization process, we design the critic agent to produce updated \textbf{critic state (i.e., the information exploration state)} including a set of \textbf{blueprints}, where each blueprint represents a key point to be covered in the final report and is accompanied by a dedicated list of targeted search queries, thereby ensuring that the information retrieval at each iteration is systematically aligned with the intended scope and coverage of the final report.
Conversely, the generator agent receives the current \textbf{critic state (i.e., the information exploration state)}, and is responsible for retrieving updated search results based on the refined search queries, and accordingly revising the generated outlines along with accompanying references, namely \textbf{generator state (i.e., the information exploitation state)}. 
Through this adversarial yet collaborative interaction, the two agents iteratively drive each other toward more comprehensive information coverage and higher-quality outline generation.

The aforementioned agentic workflow can be optimized through two broad classes of design strategies: handcrafted approaches and \textbf{automatically discovered approaches}. 
Handcrafted optimization relies on domain expert knowledge to manually engineer search heuristics and allocate computational resources across distinct components of the workflow, offering interpretability but limited scalability. 
To overcome these constraints, recent advances in meta-optimization \citep{lee2026meta} introduce an outer optimization harness that operates over the agentic workflow itself, enabling the agent to systematically explore and refine its own optimization strategies in an automated and adaptive manner—without requiring exhaustive human intervention.
Concretely, we employ Claude-Code as our primary code-generation agent and construct a meta-optimization harness around the critic agent. 
Within this harness, the \textbf{generator agent}—originally responsible for producing target outlines—is repurposed as a \textbf{scoring agent} that evaluates the critic agent's outputs and emits structured quality signals. 
This repurposing enables systematic optimization of search query generation without introducing additional model components.
Leveraging the strong code-generation capabilities of state-of-the-art agents such as Claude-Code, the framework systematically explores the space of agent configurations and automatically constructs a \textbf{policy bank}—a structured repository of reusable and composable design strategies that govern search query generation across diverse research tasks and retrieval domains. 
By drawing upon and refining the entries in this policy bank, the framework iteratively self-evolves its design strategies, progressively improving retrieval quality while reducing the need for human intervention.

We evaluate AgentDisCo on three widely
adopted deep research benchmarks, namely, DeepResearchBench \citep{du2025deepresearch}, DeepConsult \citep{DeepConsult}, and DeepResearchGym \citep{coelho2025deepresearchgym} with Gemini-2.5-Pro \citep{gemini2.5} as our base model, 
Specifically, AgentDisCo w/ Harness achieves a RACE score of \textbf{52.11} on DeepResearchBench and \textbf{6.86} on DeepConsult, surpassing leading closed-source systems such as Doubao-Research \citep{DoubaoResearch}, Claude-DeepResearch \citep{claude}, and OpenAI-DeepResearch \citep{dr}. 

However, we observe that existing benchmarks predominantly center on academic or domain-specific consulting queries, which diverge substantially from the breadth and diversity of real-world user needs. 
To bridge this gap, we introduce \textbf{GALA} (\textbf{G}eneral \textbf{A}I \textbf{L}ife \textbf{A}ssistants), a novel benchmark designed to capture authentic, everyday information-seeking behavior. 
Specifically, we collect over 10,000 highly active users from the Rednote platform\footnote{\url{https://www.xiaohongshu.com/explore}} along with their browsing and commenting histories, and devise an agentic workflow that automatically mines latent deep research interests and synthesizes personalized queries tailored to each user's individual preferences.
Compared with prior open-sourced benchmarks, the resulting queries exhibit a markedly more lifestyle-oriented character, with the dominant topics shifting from Science \& Technology and Finance \& Business toward everyday domains such as Home \& Hobbies, Fashion \& Beauty, and Travel. 
By grounding evaluation in organic user activity, GALA offers a more faithful reflection of everyday information needs and enables a more realistic assessment of deep research agents in practical deployment scenarios.
Alongside the query set, we release a standardized evaluation protocol built upon Gemini-3-Flash \citep{gemini3flash}, in which reports generated by AgentDisCo serve as reference outputs against which competing systems are scored in a pairwise manner. 
To construct competitive baselines, our human annotation team manually collects reports from the official web interfaces of Doubao-Research \citep{DoubaoResearch} and Qwen-Research \citep{QwenResearch}, as well as outputs from OpenAI o3-DeepResearch \citep{O3Research} obtained via its API. 
Experimental results demonstrate that AgentDisCo consistently outperforms these strong proprietary baselines.
Moreover, we find that AgentDisCo achieves stronger performance when relying solely on the Rednote search engine than when relying solely on Google Search, highlighting the superiority of Rednote as a source of community-grounded content for everyday information-seeking tasks.

\begin{figure}[t]
    \centering
    \includegraphics[width=1\textwidth]{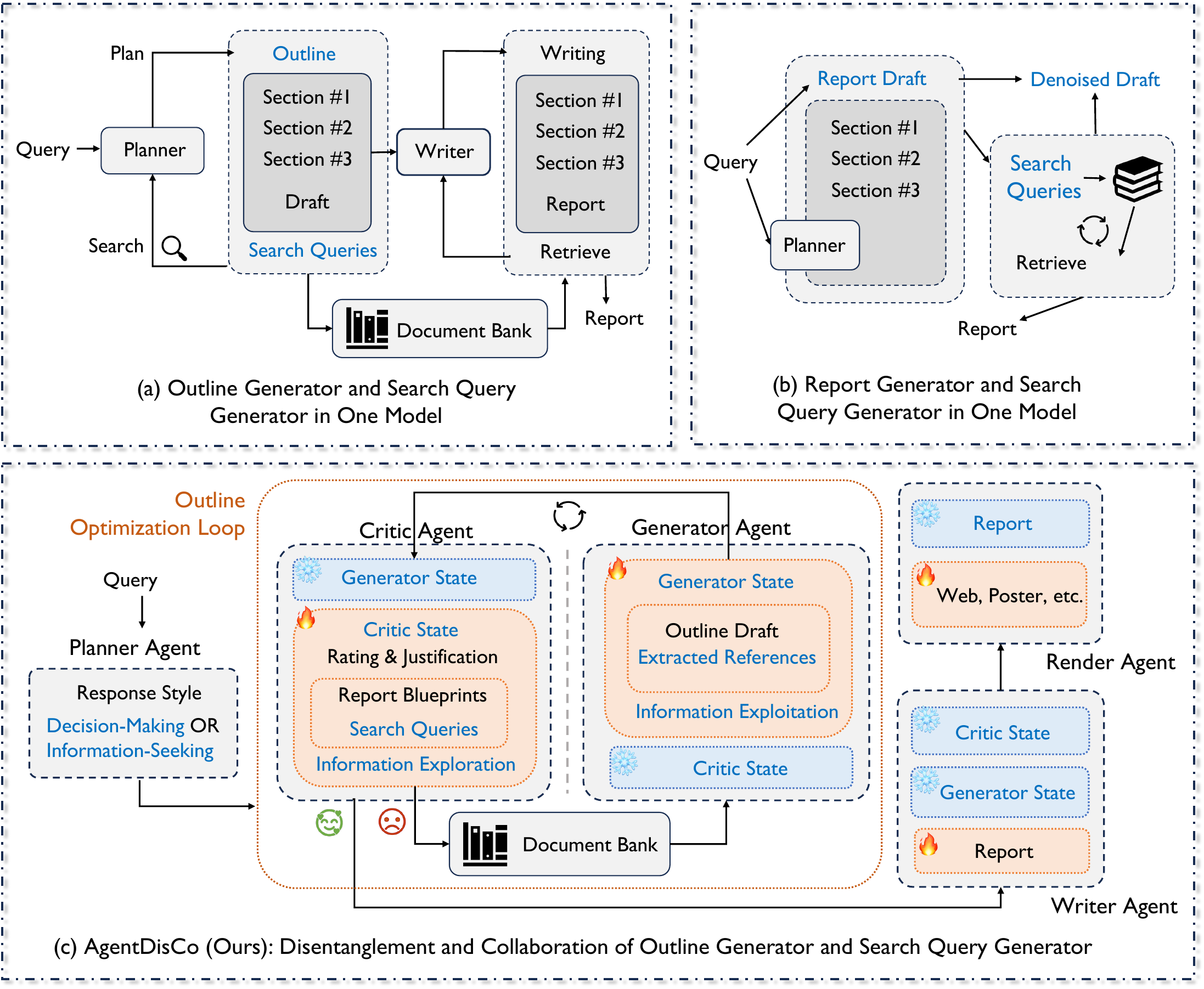}
    \caption{\textbf{Comparison of deep research paradigms.} 
    (a) the outline-iterative-optimization paradigm couples outline generation and search query formulation within a single model; (b) the report-iterative-optimization paradigm similarly entangles report generation with search query formulation; 
    (c) in contrast, \textbf{AgentDisCo} disentangles the outline generator and the search query generator into separate models, and further coordinates them through a dynamic critic-and-generator research cycle.}
    \label{fig:arch_comparison}
\end{figure}

Motivated by recent advances in AI-generated content (AIGC) and the emergence of paper-to-poster generation systems \citep{zhang2025postergen}, we further explore the integration of deep research agents with automated visual presentation. 
Recognizing that an intuitive and user-friendly interface is essential for bridging the gap between research outputs and end-user consumption, we develop a \textbf{rendering agent} that transforms structured research reports into visually rich, Rednote-style poster presentations. 
Building upon this capability, we construct a product demonstration—AutoResearch Your Interest—which automatically curates and delivers personalized deep research recommendations tailored to individual user profiles inferred from their browsing histories.

\begin{figure}[t]
    \centering
    \includegraphics[width=1\textwidth]{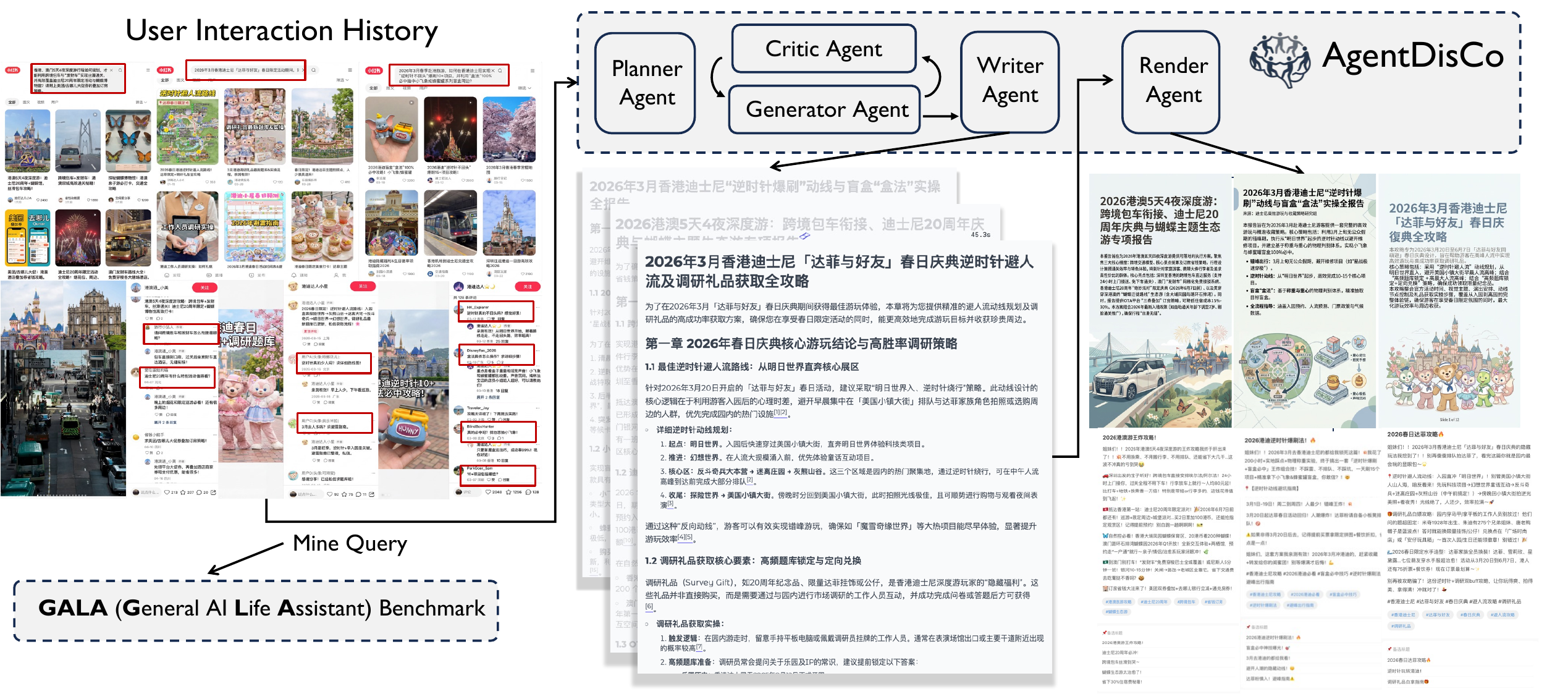}
    \caption{\textbf{Overview of the architecture and applications of AgentDisCo.} AgentDisCo spans the full pipeline from mining latent deep research queries in user interaction histories to producing structured reports and rendering visually rich posters. This end-to-end design realizes the vision of ``AutoResearch Your Interest''—automatically tracking evolving user interests and delivering personalized deep research recommendations tailored to individual user profiles.}
    \label{fig:arch_application}
\end{figure}

\textbf{Contributions.}
Our main contributions can be summarized as follows.
\begin{itemize}[leftmargin=*, itemsep=2pt, topsep=2pt, parsep=0pt]
    \item \textbf{A novel disentangled and collaborative agentic architecture.} As illustrated in Figure~\ref{fig:arch_comparison}, we introduce \textbf{AgentDisCo}, which formulates deep research as an adversarial optimization problem between information exploration and exploitation, decoupling outline generation from search query formulation and coordinating them through a dynamic critic-and-generator cycle.
    
    \item \textbf{A meta-optimization harness for self-evolving search.} We construct a meta-optimization harness over the adversarial optimization loop, in which the generator agent—originally responsible for producing target outlines—is repurposed as a scoring agent that evaluates the critic agent's outputs and emits structured quality signals. This design enables the systematic and automatic optimization of search query generation without introducing additional model components.

    \item \textbf{A new benchmark for lifestyle deep research needs.} Observing that existing benchmarks predominantly focus on academic or domain-specific consulting queries—diverging substantially from the breadth and diversity of real-world user needs—we introduce \textbf{GALA}, a benchmark that captures authentic, lifestyle-oriented information-seeking behavior mined from organic user activity.

    \item \textbf{An open-sourced deep research system with multi-modal render agent.} Recognizing that an intuitive interface is essential for bridging research outputs and end-user consumption, we develop a \textbf{rendering agent} that transforms structured reports into visually rich, Rednote-style poster presentations. As depicted in Figure~\ref{fig:arch_application}, AgentDisCo thus spans the full pipeline—from deep research interest mining (which underpins the GALA benchmark) to the generation of reports and posters. To support and accelerate future research on open-ended deep research, we publicly release our benchmark, code, demo, and evaluation harness.
\end{itemize}

\section{AgentDisCo: A Disentangled and Collaborative Agentic Architecture}
\label{sec:arch}

\subsection{System Overview and Design Philosophy}
\label{sec:overview}
We consider open-ended deep research questions without ground-truth answers.
Given such a question $q$, the agentic system must perform \textbf{information exploration} (searching relevant evidence) and \textbf{information exploitation} (synthesizing the evidence into a structured report).
We argue that iterative optimization should occur on \textbf{intermediate outlines} rather than final reports, and we further disentangle exploration and exploitation into two specialized but interacting agents — a \textbf{critic agent} $\pi^c$ and a \textbf{generator agent} $\pi^g$ — whose adversarial yet collaborative interplay drives convergence toward a comprehensive, well-grounded outline.

We formalize this disentangled and collaborative interaction as a dual-agent cooperative MDP (Markov Decision Process):
\begin{equation}
\mathcal{M} = \langle \mathcal{S}^\mathrm{c}, \mathcal{S}^\mathrm{g}, \mathcal{A}^\mathrm{c}, \mathcal{A}^\mathrm{g}, \mathcal{P}, \mathcal{R}, T \rangle,  
\end{equation}
where the joint environment state at timestep $t$ is the pair $\mathrm{s}_t = (\mathrm{s}^\mathrm{c}_t, \mathrm{s}^\mathrm{g}_t)$.
The \textbf{generator state (i.e., the information-exploitation state)}, denoted $\mathrm{s}^\mathrm{g}_t = (\mathrm{O}_t, \mathrm{R}_t)$, consists of the current outline $\mathrm{O}_t$ and the set of references $\mathrm{R}_t$ attached to it.
The \textbf{critic state (i.e., the information-exploration state)}, denoted $\mathrm{s}^\mathrm{c}_t = (\mathrm{B}_t, \mathrm{Q}_t)$, comprises a set of blueprints $\mathrm{B}_t$ together with their associated search queries $\mathrm{Q}_t$.
Each blueprint in $\mathrm{B}_t$ specifies a key point that the final report should cover, and each such key point is paired with a list of targeted search queries in $\mathrm{Q}_t$ dedicated to filling its information gap.
Together, the blueprints align retrieval with the intended scope 
and coverage of the report.

Given a user query $q$, the system initializes the generator state as $\mathrm{s}^\mathrm{g}_0 = (\varnothing, \varnothing)$, indicating an empty outline and an empty document pool.
Conditioned solely on $q$, the critic agent then performs a coarse decomposition and proposes an initial blueprint set $\mathrm{s}^\mathrm{c}_0 \sim \pi^\mathrm{c}(\cdot \mid q, \varnothing, \varnothing)$, which seeds the subsequent iterative refinement.
At each iteration $t$, the two agents act sequentially rather than simultaneously, with the critic moving first.
At the critic's step, conditioned on the question $q$, the previous generator state $\mathrm{s}^\mathrm{g}_{t-1}$ and its own previous state $\mathrm{s}^\mathrm{c}_{t-1}$, the critic agent assesses the completeness and quality of the current outline and produces an updated blueprint set:
$\mathrm{s}^\mathrm{c}_t \sim \pi^\mathrm{c}(\cdot \mid q, \mathrm{s}^\mathrm{g}_{t-1}, \mathrm{s}^\mathrm{c}_{t-1})$.
That is, the critic's action is to instantiate the next exploration state.
At the generator's step, conditioned on the freshly updated critic state $\mathrm{s}^\mathrm{c}_{t}$ together with its previous state $\mathrm{s}^\mathrm{g}_{t-1}$, the generator agent executes the queries prescribed by each blueprint, retrieves new evidence via the search tool, and accordingly revises both the outline and its references: $\mathrm{s}^\mathrm{g}_t \sim \pi^\mathrm{g}(\cdot \mid q, \mathrm{s}^\mathrm{g}_{t-1}, \mathrm{s}^\mathrm{c}_{t})$.
The generator's action thus realizes the exploitation counterpart of the critic's exploration.
As for environment dynamics, given the actions of both agents, the joint state transition is deterministic: each agent's output directly instantiates the corresponding component of the next joint state, i.e., $\mathrm{s}^\mathrm{c}_{t+1}=\mathrm{a}^\mathrm{c}_t$ and $\mathrm{s}^\mathrm{g}_{t+1}=\mathrm{a}^\mathrm{g}_t$.
Consequently, all stochasticity of the trajectory originates from the agents' policies themselves.

Under the above sequential protocol, an interaction trajectory of AgentDisCo unfolds as the alternating sequence $\tau = \{\, q,\, \mathrm{s}^\mathrm{c}_0,\, \mathrm{s}^\mathrm{g}_0,\, \mathrm{s}^\mathrm{c}_1,\, \mathrm{s}^\mathrm{g}_1,\, \cdots \,\}$, in which the critic and the generator alternately update their respective state components. 
The likelihood of sampling $\tau$ admits the following equivalent factorizations:
\begin{equation}
\label{eqn:dual}
\begin{aligned}
p(\tau)&=\prod_{t=0}^{T-1}\underbrace{\pi^{\mathrm{c}}\bigl(\mathrm{s}^{\mathrm{c}}_{t+1}\mid q, \mathrm{s}^{\mathrm{g}}_{t}, \mathrm{s}^{\mathrm{c}}_{t}\bigr)}_{\textcolor[RGB]{39,97,208}{\text{(1) Critic agent policy}}}\underbrace{\pi^{\mathrm{g}}\bigl(\mathrm{s}^{\mathrm{g}}_{t+1}\mid q, \mathrm{s}^{\mathrm{g}}_{t}, \mathrm{s}^{\mathrm{c}}_{t+1}\bigr)}_{\textcolor[RGB]{231,72,49}{\text{(2) Generator agent policy}}}\\
&=\prod_{t=0}^{T-1}\underbrace{\pi^{\mathrm{c}}\bigl(\mathrm{a}^{\mathrm{c}}_{t}\mid q, \mathrm{s}^{\mathrm{g}}_{t}, \mathrm{s}^{\mathrm{c}}_{t}\bigr)}_{\textcolor[RGB]{39,97,208}{\text{(1) Critic agent policy}}}
\underbrace{\mathbb{I}\bigl[\ \mathrm{s}^{\mathrm{c}}_{t+1} = \mathrm{a}^{\mathrm{c}}_{t} \bigr]\  \mathcal{P}^{\mathrm{c}}\bigl(\mathrm{s}^{\mathrm{g}}_{t+1}\mid q, \mathrm{s}^{\mathrm{g}}_{t}, \mathrm{a}^{\mathrm{c}}_{t}\bigr)}_{\textcolor[RGB]{231,72,49}{\text{(2) Environment dynamic of critic agent}}}\\
&=\prod_{t=0}^{T-1}\underbrace{\pi^{\mathrm{g}}\bigl(\mathrm{a}^{\mathrm{g}}_{t}\mid q, \mathrm{s}^{\mathrm{g}}_{t}, \mathrm{s}^{\mathrm{c}}_{t+1}\bigr)}_{\textcolor[RGB]{231,72,49}{\text{(1) Generator agent policy}}}
\underbrace{\mathbb{I}\bigl[\ \mathrm{s}^{\mathrm{g}}_{t+1} = \mathrm{a}^{\mathrm{g}}_{t} \bigr]\ \mathcal{P}^{\mathrm{g}}\bigl(\mathrm{s}^{\mathrm{c}}_{t+2}\mid q, \mathrm{a}^{\mathrm{g}}_{t}, \mathrm{s}^{\mathrm{c}}_{t+1}\bigr)}_{\textcolor[RGB]{39,97,208}{\text{(2) Environment dynamic of generator agent}}}\\
\end{aligned}
\end{equation}
where $T$ denotes the maximum number of optimization rounds and $\mathbb{I}(\cdot)$ is the indicator function.
The first line presents the most compact policy-only factorization, while the latter two further decouple each agent's stochastic decision from its (degenerate) environment transition, thereby making explicit the loci at which learning signals can be injected.

Although both agents are optimized toward a common objective, their per-round roles are functionally adversarial, jointly forming a minimax-style yet cooperative loop.
The critic agent $\pi^c$ adversarially probes the generator's current outline, surfacing missing evidence and uncovered key points through newly proposed blueprints — thereby pushing exploration outward.
The generator agent $\pi^g$ defensively expands and grounds the outline using the freshly retrieved evidence — thereby consolidating exploitation inward.
This adversarial-yet-aligned interaction progressively refines the outline along the axes of coverage, factual grounding, and structural coherence.
To reconcile the two locally adversarial roles under a single global objective, both agents share a common cooperative reward that quantifies the quality of the updated outline for the subsequent writer agent and render agent.

Below, we describe each component of AgentDisCo in detail (as depicted in Figure~\ref{fig:arch_comparison}), including the aforementioned (outline) critic and (outline) generator agents.

\begin{figure}[t]
    \centering
    \includegraphics[width=1\textwidth]{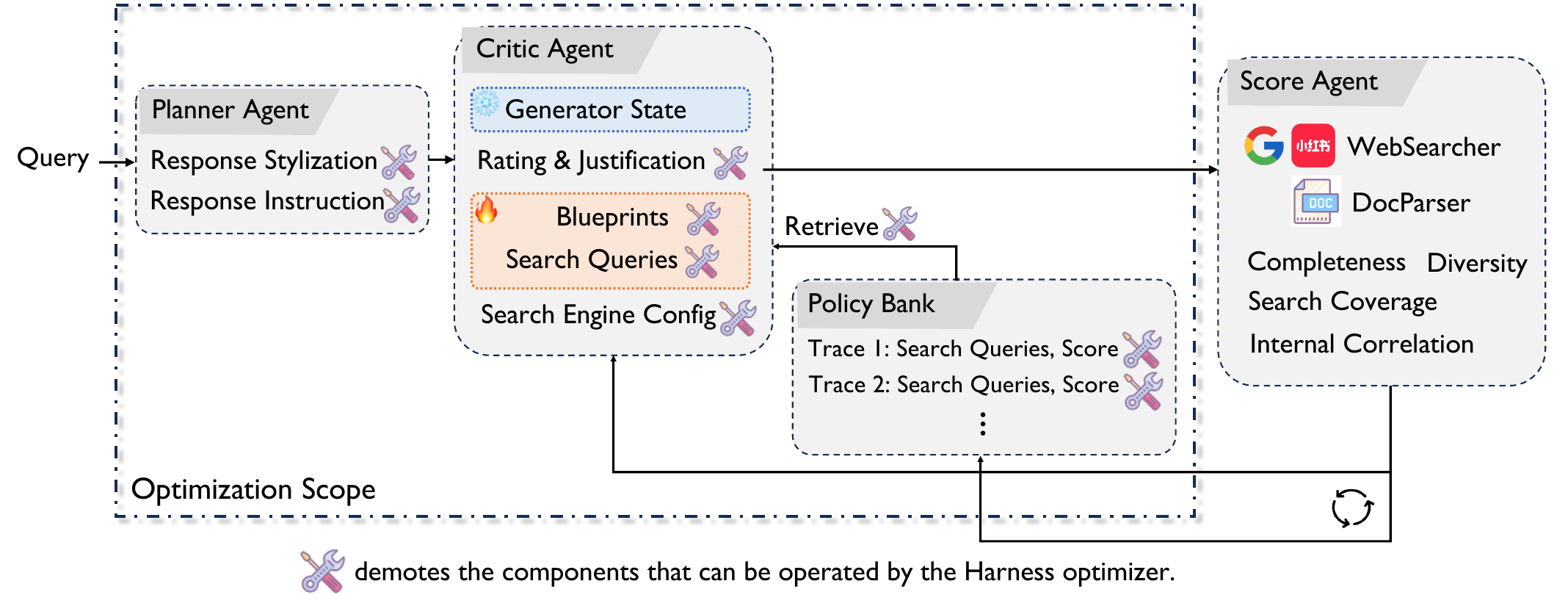}
    \caption{\textbf{Overview of the harness optimization in AgentDisCo.} AgentDisCo can automatically discover design strategies by constructing a meta-optimization harness around the adversarial optimization loop. Specifically, the generator agent—originally tasked with producing target outlines—is repurposed as a scoring agent that evaluates the critic agent's outputs and generates quality signals, thereby enabling systematic optimization of the search queries.}
    \label{fig:agent_harness}
\end{figure}

\subsection{Planner Agent}
\label{sec:planner}

\textbf{Handcrafted Design.}
To accommodate the heterogeneous nature of deep research queries encountered on the industrial platform, we introduce a planner agent that classifies each incoming query into one of two top-level categories, each further subdivided into fine-grained intents, namely \textbf{information seeking} category (including fact query, status \& progress, news \& information, deep exploration, and resource locating) and \textbf{decision making} category (comparison \& selection, recommendations \& suggestions, how-to guide, travel planning, and purchase decision).
Beyond the categorical label, the planner also infers an expected response style tailored to the predicted intent, which subsequently conditions the downstream critic and generator agents. 
We instantiate the planner with Gemini-2.5-Pro \citep{gemini2.5} as the backbone, and provide the full prompt template in Appendix~\ref{app:planner_prompt}.

\textbf{Harness Optimization.} During harness optimization, we grant the planner agent access to generate instructions for the subsequent agentic workflows. 
Concretely, we observe that our code-generation agent (i.e., the Claude-Code agent) specifies instructions such as tips for formulating search queries and selecting hyperparameters for the search engines, which are then passed along to the downstream (outline) critic and generator agents. 
Detailed prompts are provided in Appendix~\ref{app:harness_prompt}.

Formally, we introduce the notation $\pi_{\text{Planner}}(\cdot)$ to denote the planner agent, whose function can be written as $\mathrm{P} \sim \pi_{\text{Planner}}(q)$.
In the handcrafted design, $\mathrm{P}$ represents the intent type and response style conditioned on the input query $q$, whereas during harness optimization, $\mathrm{P}$ additionally includes specified instructions.
For ease of representation, we omit $\mathrm{P}$, since it can be regarded as a complementary explanation attached to the query and thus integrated into $q$, namely $q = [\mathrm{P};\, q]$.

\subsection{(Outline) Critic Agent and (Outline) Generator Agent}
\label{sec:critic}
\textbf{Handcrafted Design.} 
As described in Section~\ref{sec:overview}, the core of AgentDisCo lies in the dual optimization between the critic and generator agents. 
Eq.~(\ref{eqn:dual}) reveals that, although both agents assume functionally adversarial roles in each round, they jointly constitute a minimax-style yet cooperative loop: both are optimized toward a shared objective, namely, the production of a high-quality outline that can attain a sufficiently high score under the critic's evaluation.
The resulting artifact, denoted as $\mathrm{O}_t$ and $\mathrm{R}_t$, serves as the foundation for the subsequent writing and rendering stages.

To assess outline quality, our handcrafted design employs the critic agent to produce both a numerical score and an accompanying justification for each candidate outline. The optimization loop is further governed by three control parameters: an exit threshold that determines when the outline is deemed acceptable by the critic, a minimum number of optimization rounds to ensure sufficient refinement, and a maximum number of rounds to bound the overall computational cost.
Formally, the reward function for $(\mathrm{O}_t, \mathrm{R}_t)$ can be expressed as $\mathcal{R}(\mathrm{s}_t^\mathrm{g}, \mathrm{a}_t^\mathrm{g})$, where $\mathrm{a}_t^\mathrm{g}$ denotes the generator's action, i.e., the produced outline $\mathrm{O}_t$ and its rendering $\mathrm{R}_t$. 
At iteration $t+1$, the critic agent receives $(\mathrm{O}_t, \mathrm{R}_t)$ as input and outputs an evaluation score that constitutes the reward signal. 
Accordingly, the reward is sampled as $\mathrm{r}_t \sim \pi^\mathrm{c}(\mathrm{a}^\mathrm{c}_{t+1} \mid \mathrm{s}^\mathrm{c}_{t+1})$, with the critic's state defined as $\mathrm{s}^\mathrm{c}_{t+1} = \mathrm{a}_t^\mathrm{g}$, thereby coupling the generator's output directly to the critic's input.

A persistent challenge in multi-round outline optimization is reference management: as the optimization unfolds, the agent must continually decide which documents retrieved in earlier rounds are worth retaining for downstream use and which can be safely discarded. 
Naive strategies risk either prematurely dropping high-quality evidence or overloading the context window with stale and redundant content—both of which degrade the quality of the final outline.
To address this, we introduce the \textbf{document bank}, a lightweight recorder and tracker that maintains a persistent view of retrieved references across rounds. 
Once the critic agent produces a set of search queries, the document bank parses each retrieved document into fine-grained evidence snippets, scores documents in parallel for relevance, summarizes their content, and extracts key evidence triples; low-scoring documents are filtered out before reaching the generator agent. 
In this way, the document bank not only compresses raw search results into a structured, citation-ready memory, but also shields the generator agent from contextual noise and redundancy.

To ensure stable progression and prevent information loss across rounds, each new optimization round enforces the following continuity constraints.
Firstly, all documents contained in the reference set (i.e., $\mathrm{R}_t$ in the outline $\mathrm{O}_t$) are carried over as input to the next round (i.e., $t+1$), guaranteeing that previously validated evidence remains accessible to both agents. 
The document bank correspondingly updates its document indices to align with the new round.
Secondly, while outlines and search queries are fully re-generated at each round, \textbf{the underlying blueprint is only permitted to be modified or expanded}; deletion of existing key points is discouraged, thereby preserving the structural backbone established in earlier rounds.
Moreover, each invocation of the critic agent (for refining search queries and blueprints) and the generator agent (for refining outlines) is explicitly conditioned on the concrete content produced in the preceding iteration, ensuring that optimization proceeds incrementally rather than restarting from scratch.

\textbf{Harness Optimization.} 
Our central insight is that LLMs excel at extracting and summarizing information, but are comparatively weaker at generating effective search queries across heterogeneous retrieval sources and at organizing coherent outlines. 
In this paper, we therefore focus on harness-based optimization for search query generation, leaving outline organization to future work.
Concretely, we develop a harness that optimizes the (outline) critic agent, whose responsibility is to formulate effective search queries and retrieve high-quality information from web-scale sources. 
As illustrated in Figure~\ref{fig:agent_harness}, we further repurpose the generator agent as a score agent that analyzes the returned search results and provides feedback signals to guide the critic's query refinement. 
In practice, since the search queries are organized under a set of blueprints, evaluating individual results in isolation is insufficient. 
We therefore design several criteria—namely completeness, diversity, search coverage, and internal correlation—and apply them to each search result, aggregating the per-result scores into statistics and distributions over the entire result set. 
These aggregated signals provide the critic with a holistic view of retrieval quality, enabling more targeted query refinement in subsequent rounds. 
The detailed prompt used by the score agent is provided in the Appendix~\ref{app:harness_prompt}.

\begin{figure}[t]
    \centering
    \includegraphics[width=1\textwidth]{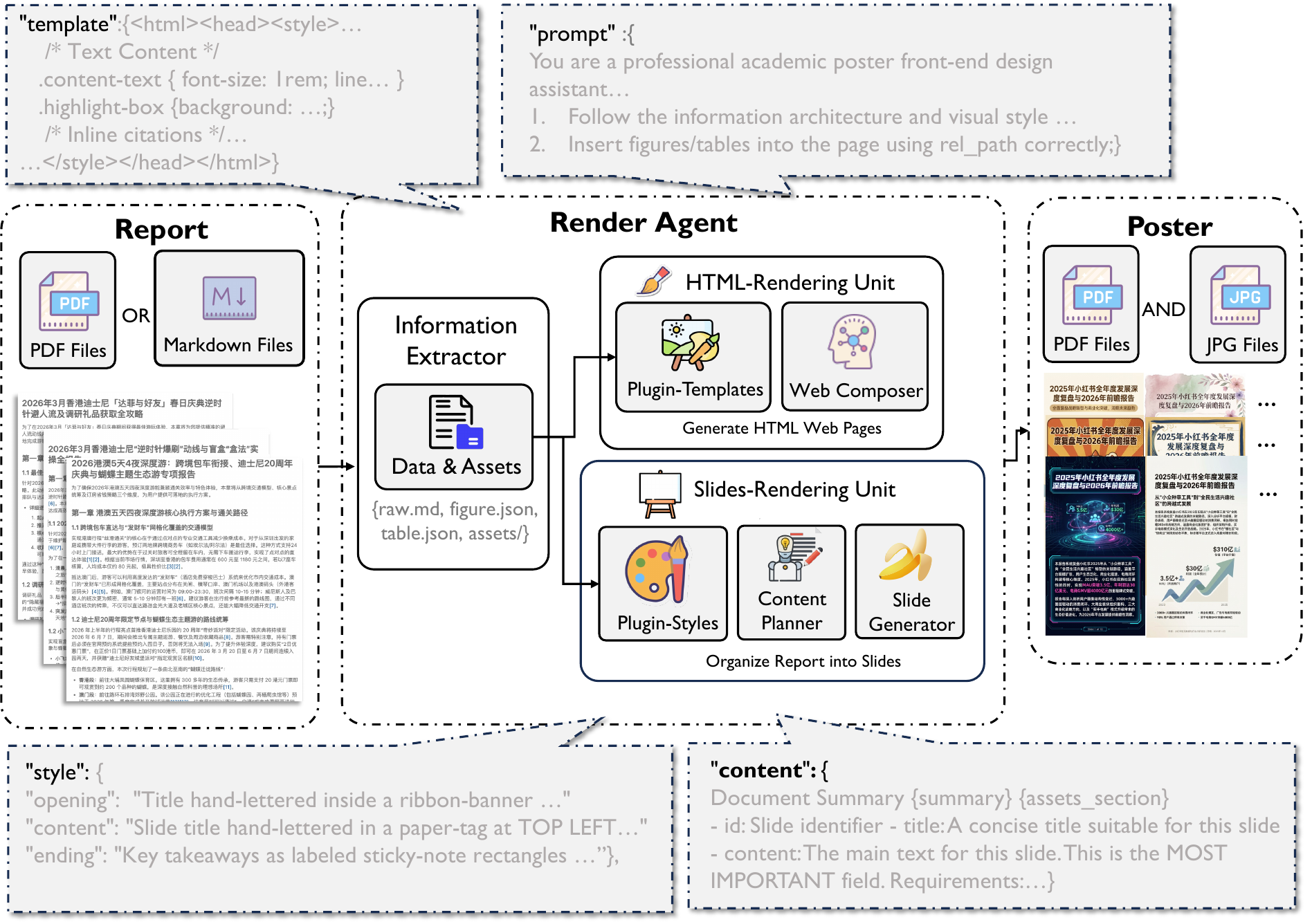}
    \caption{\textbf{Overview of the render agent in AgentDisCo.} Our render agent accepts as input a report in either PDF or Markdown format. It first extracts the salient features and structural elements from the report, and then reorganizes the content into one of two presentation modalities: an HTML-based layout or a slide-style layout. Notably, both modalities support pluggable templates and styling components, enabling flexible visual customization. The final output is rendered as a PDF document or a sequence of images, depending on the chosen modality.}
    \label{fig:agent_render}
\end{figure}

A particularly noteworthy emergent behavior arises during harness optimization: prompted only by a minimal cue—``you are allowed to store and retrieve traces to evolve''—the coding agent (i.e., the Claude-Code agent) autonomously builds a policy bank that records and reuses relevant historical traces, including critic states, generator states, and the aforementioned criterion scores, to guide subsequent optimization.
Formally, the likelihood of sampling $\tau$ can be formulated as follows:
\begin{equation}
\label{eqn:sampletau}
p(\tau)=\prod_{t=0}^{T-1}
\underbrace{\mu^\mathrm{r}\bigl(\mathrm{m}_t\mid q, \mathrm{s}^{\mathrm{g}}_{t}, \mathrm{s}^{\mathrm{c}}_{t},\mathrm{M}_t\bigr)}_{{\text{(1) Retrieve from policy bank}}} 
\underbrace{\pi^{\mathrm{c}}\bigl(\mathrm{a}^{\mathrm{c}}_{t}\mid q, \mathrm{s}^{\mathrm{g}}_{t}, \mathrm{s}^{\mathrm{c}}_{t},\mathrm{m}_t\bigr)}_{\textcolor[RGB]{39,97,208}{\text{(2) Critic agent policy}}}
\underbrace{\mu^\mathrm{w}\bigl(\mathrm{M}_{t+1}|\cdot\bigr)}_{{\text{(3) Update policy bank}}} 
\underbrace{\mathbb{I}\bigl[\ \mathrm{s}^{\mathrm{c}}_{t+1} = \mathrm{a}^{\mathrm{c}}_{t} \bigr]\  \mathcal{P}^{\mathrm{c}}\bigl(\mathrm{s}^{\mathrm{g}}_{t+1}\mid q, \mathrm{s}^{\mathrm{g}}_{t}, \mathrm{a}^{\mathrm{c}}_{t}\bigr)}_{\textcolor[RGB]{231,72,49}{\text{(4) Environment dynamic of critic agent}}}
\end{equation}
Here, $\mathrm{m}_t$ denote the trace(s) sampled at step $t$ from the policy bank $\mathrm{M}_t$.
The retrieval process is formulated as $\mathrm{m}_t \sim \mu^\mathrm{r}\bigl(\cdot \mid q, \mathrm{s}^{\mathrm{g}}_{t}, \mathrm{s}^{\mathrm{c}}_{t},\mathrm{M}_t\bigr)$, where $\mu^\mathrm{r}$ denotes the retrieval function.
In practice, $\mu^\mathrm{r}$ is instantiated as a simple BM25-based retriever \citep{robertson2009probabilistic}, which is autonomously implemented by the Claude-Code agent during harness optimization.
After each step, the newly produced trace is written back into the policy bank, yielding the updated bank $\mathrm{M}_{t+1}\sim \mu^\mathrm{w}\bigl(\cdot|q, \mathrm{s}^{\mathrm{g}}_{t}, \mathrm{s}^{\mathrm{c}}_{t},\mathrm{m}_t,\mathrm{a}^{\mathrm{c}}_{t},\mathrm{M}_t\bigr)$, where $\mu^\mathrm{w}$ is a lightweight writing function.
Together, $\mu^\mathrm{r}$ and $\mu^\mathrm{w}$ constitute a simple yet effective read-write interface that enables the policy bank to evolve continuously throughout the optimization process.

\subsection{Writer Agent}
\label{sec:writer}
As introduced in Section~\ref{sec:critic}, the document bank effectively filters out irrelevant content, thereby alleviating the burden on the model's attentional capacity. Since the generated outline is inherently structured, it is straightforward to partition the outline together with its associated references into a sequence of self-contained chunks, following the strategy of \citet{li2025webweaverstructuringwebscaleevidence}. 
This decomposition reduces the complex task of long-context writing into a series of manageable, attention-focused subtasks, each operating on only the relevant evidence.
Because each document has already been assigned a unique index within the outline, the relevant evidence for any given section can be retrieved directly from the document bank. 
The composition of each section is therefore not a single monolithic action, but rather a deliberate intra-sectional reasoning cycle: at each step, the writer conditions on the previously generated chunks and continues the narrative in a coherent, context-aware manner. 
This internal monologue is critical for moving beyond shallow summarization toward genuine synthesis across evidence.
Finally, the system outputs a Markdown-formatted report in which every cited reference is accompanied by its corresponding URL, ensuring full source traceability.
Notably, the writer agent is additionally conditioned on the response style produced by the planner agent (Section~\ref{sec:planner}), ensuring that the generated report remains coherent with the input query and faithfully aligned with the user's intent.

As discussed in Section~\ref{sec:critic}, our key premise is that extracting and summarizing information does not constitute the primary bottleneck for LLMs, in contrast to the more challenging task of generating effective search queries across heterogeneous retrieval sources. 
Accordingly, this paper focuses on harness-based optimization for search query generation, leaving the optimization of the writer agent to future work.

\begin{figure}[t]
    \centering
    \includegraphics[width=1\textwidth]{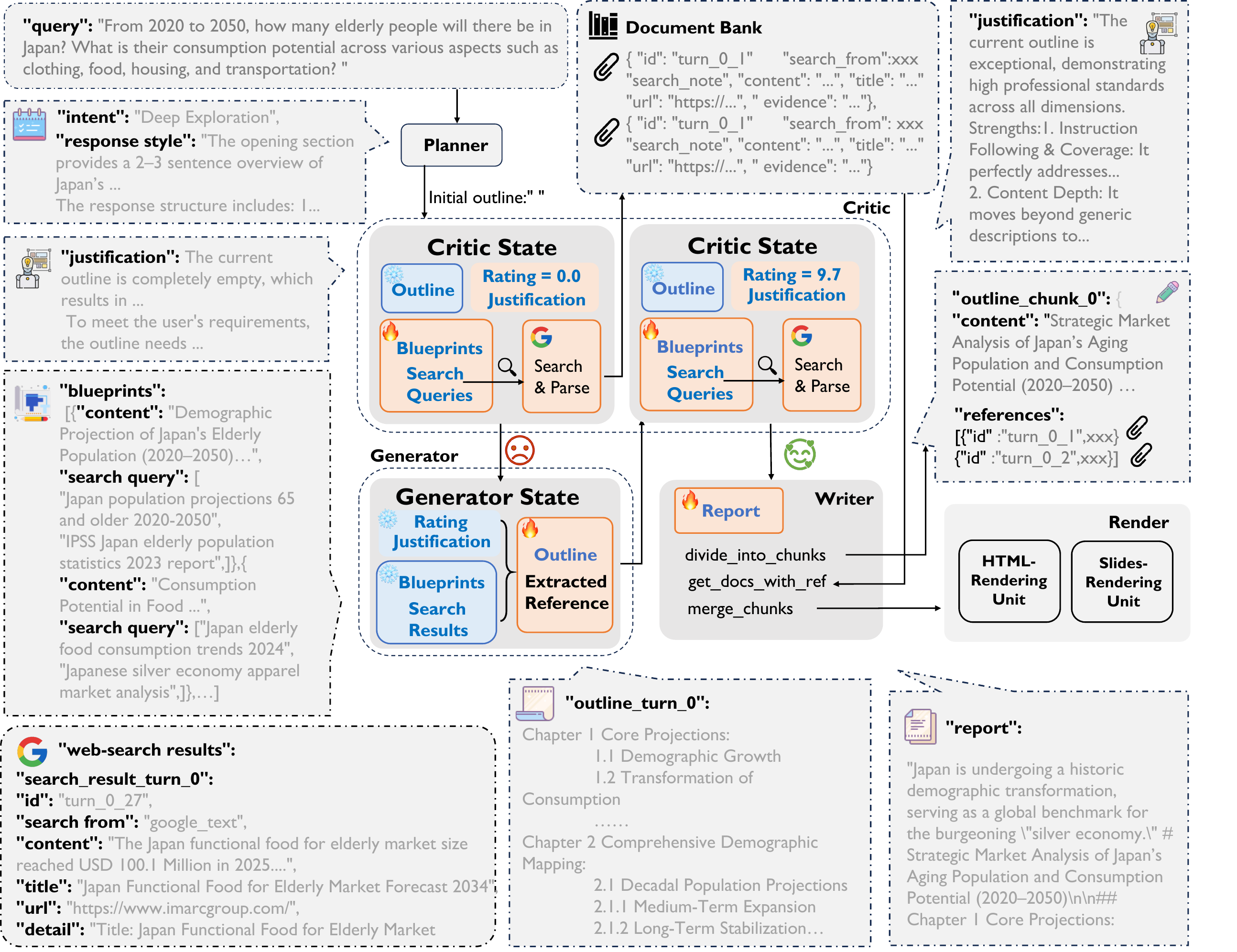}
    \caption{\textbf{A showcase of AgentDisCo.} The figure illustrates the end-to-end processing pipeline of an input query: starting from the planner agent, proceeding through the iterative optimization loop between the (outline) critic agent and the (outline) generator agent, and finally passing to the writer agent and the render agent.}
    \label{fig:agent_log}
\end{figure}

\subsection{Render Agent}
\label{sec:render}
Considering that a clear, intuitive, and visually engaging interface is essential for bridging the gap between raw research outputs and end-user consumption, we develop a render agent that transforms structured research reports into visually rich presentations—ranging from RedNote-style posters to slide decks and HTML pages—thereby allowing users to readily digest, share, and act upon the generated content.
Existing open-source render agents~\citep{p2p,postergen,paper2page,paper2poster,glyphbanana} rely heavily on complex, tightly coupled pipelines that target a single output modality, making it difficult to accommodate heterogeneous user preferences over presentation forms. 
Moreover, their intricate iterative design and rigid workflow orchestration limit the extensibility of packaging multimodal pipelines as reusable tools or skills, while incurring additional deployment and debugging overhead. 
In contrast, we present a clean and flexible render agent that generalizes across diverse rendering tasks within a unified framework.

As depicted in Figure~\ref{fig:agent_render}, our render agent incorporates an information extractor to extract key information points from the report. To enhance its performance, we also feed the blueprints and response style into the information extractor as auxiliary inputs.
To ensure flexibility in use, we construct a webpage template set $\mathcal{T}^\mathrm{w}$ and a slide template set $\mathcal{T}^\mathrm{s}$. 
Given an input document, the information extractor first produces structured multimodal assets conditioned on either $\mathcal{T}^\mathrm{w}$ or $\mathcal{T}^\mathrm{s}$.
We then apply Gemini-2.5-Pro \citep{gemini2.5} as the web composer to generate webpages in HTML format, or Gemini-3-Pro-Image \citep{gemini-3-pro-image} as the slide generator to produce multiple images for constructing posters or slides.
In addition, we provide an option to generate Rednote-style posters, in which the textual content is produced by Gemini-2.5-Pro \citep{gemini2.5} acting as the content planner.
We present multiple showcases of our rendered posters in the Appendix~\ref{app:showcase_render}, along with interactive demos.

\subsection{A Running Example}
\label{sec:example}
To illustrate the end-to-end workflow of AgentDisCo, we present a real execution trace in Figure~\ref{fig:agent_log} for the example query: ``From 2020 to 2050, how many elderly people will there be in Japan, and what is their consumption potential across various aspects such as clothing, food, housing, and transportation?''.

First, the planner agent generates a response style based on the input query, which is then passed, together with the query, into the optimization loop between the outline critic agent and the outline generator agent.
At round $0$, the critic agent observes an empty outline and assigns a rating of $0$. 
At round $1$, the critic agent evaluates the draft and identifies that the housing section is underdeveloped. 
Accordingly, it emits an updated set of blueprints with refined search queries specifically targeting elderly housing consumption. 
The generator then revises the outline in response, while the document bank preserves the valid citations accumulated from turn $0$, ensuring incremental knowledge accumulation across turns. 
This trace illustrates how the outline critic and generator agents co-evolve through the exchange of blueprints, rather than one dominating the other.
Finally, the outline that receives the highest score is forwarded to the downstream agents: the writer agent composes the report by sequentially elaborating each section of the outline, drawing on the associated references retrieved from the document bank, and the render agent subsequently transforms the report into the desired presentation format.

\section{GALA: A Benchmark for General AI Life Assistants}
\label{sec:gaia}
\subsection{Data Collection and Synthesis on Deep Research Queries}

\begin{figure}[t]
    \centering
    \includegraphics[width=1\textwidth]{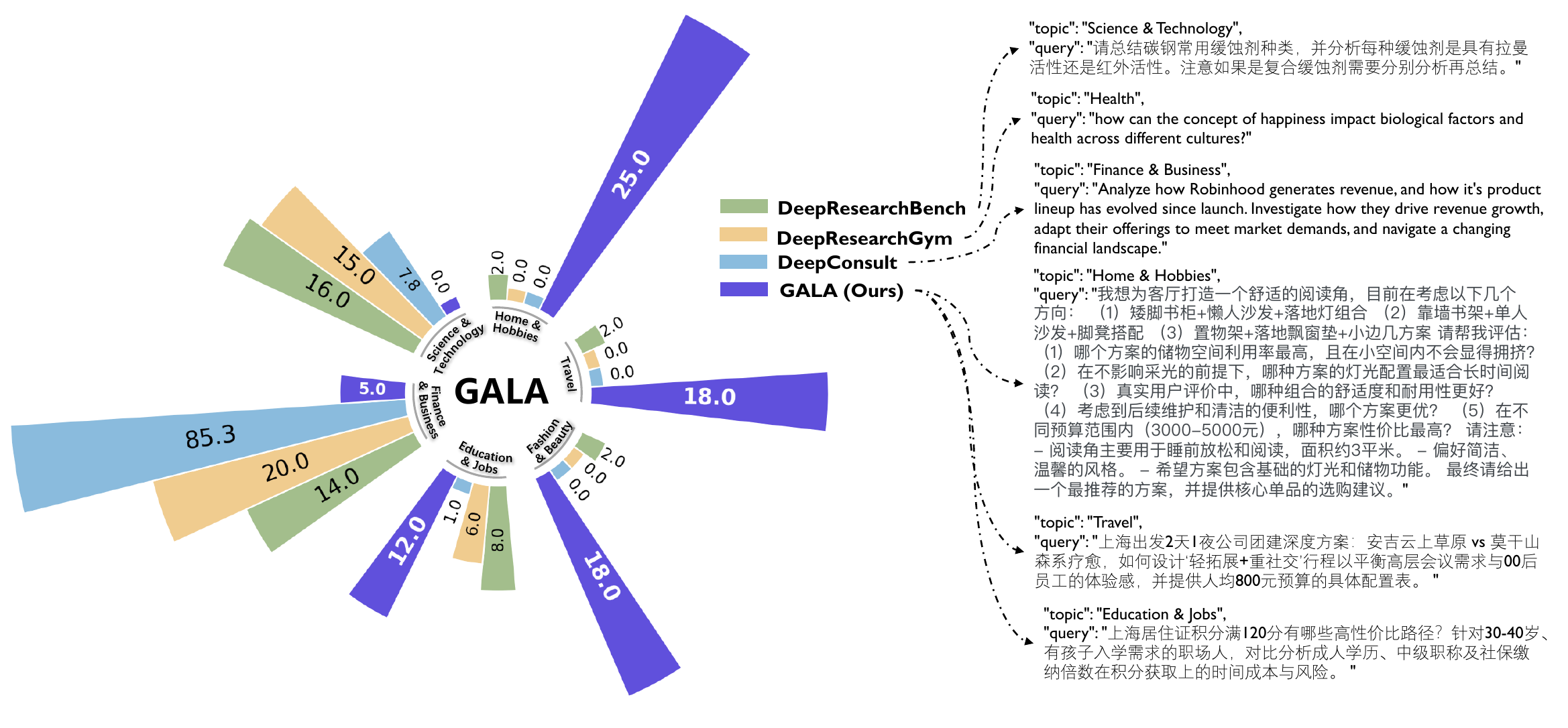}
    \caption{\textbf{Comparison between our proposed GALA benchmark with existing benchmarks DeepResearchBench, DeepResearchGym, DeepConsult.} .}
    \label{fig:agent_gala}
\end{figure}

Early deep research agents primarily focused on isolated tasks such as question answering and translation, and later advanced through tool integration to enable autonomous information retrieval and synthesis. 
To evaluate such systems, a variety of deep research benchmarks~\citep{DeepConsult,coelho2025deepresearchgym,du2025deepresearch} have been developed, with their queries typically sourced from in-house datasets of raw user interactions with web-search-enabled LLM chatbots. 
However, these collected queries predominantly fall into a narrow set of categories, such as scientific reports or consulting-style solutions, and thus fail to reflect the diversity of real-world information needs.
In contrast, Rednote has emerged as a popular search platform for daily-life information needs, covering a broad spectrum of topics ranging from travel and lifestyle to consumption and entertainment. 
Leveraging the massive user-interaction and content data on the Rednote platform, we collect the top 10{,}000 highly active users together with their historical interactions---including clicks, comments, and browsing records---to mine their latent deep research needs. 
For example, as shown in Figure~\ref{fig:arch_application}, if a user repeatedly browses or comments on notes related to ``Hong Kong Disneyland Duffy and Friends Springtime Festival'', we synthesize a corresponding deep research query that captures the user's underlying intent, such as ``During the `Duffy and Friends' Springtime limited-edition event at Hong Kong Disneyland in March 2026, how can one plan a counter-clockwise touring route to avoid crowds, and obtain the latest question bank and practical procedures for the staff survey gifts?''
Concretely, we employ a strong LLM (i.e., Gemini-3-Flash~\citep{gemini3flash}) to extract latent deep research interests from user interactions and reformulate them into well-structured deep research queries. Detailed prompts are provided in Appendix~\ref{app:query_miner}.

To ensure rigor and high quality, we further apply an inspection pipeline that distills 100 high-quality queries from an initial pool of 260{,}000 generated candidates. 
The pipeline combines automated LLM-based screening with human verification, together with an optional difficulty-expansion step in between. 
Specifically, we form a review committee powered by Gemini-3-Pro~\citep{gemini-3-pro} to automatically evaluate each query along the following criteria: 
(i) naturalness and clarity: whether the query is fluent, unambiguous, and faithfully reflects a plausible user intent;
(ii) indispensability of Rednote-specific knowledge: whether answering the query genuinely relies on Rednote's user-generated content or community insights, rather than being trivially solvable via generic web search, while ensuring that no user privacy is compromised; 
(iii) real-world plausibility: whether the query type aligns with realistic information-seeking behaviors observed in daily life, rather than representing a contrived or synthetic use case.
Queries that pass this automatic review are then forwarded to human annotators for final verification, ensuring that the resulting benchmark is both authentic and challenging.

\subsection{Statistics and Comparisons with Existing Benchmarks}
Our GALA benchmark consists of 100 carefully curated deep research queries.
Following the taxonomy proposed in DeepResearchBench \citep{du2025deepresearch}, each query is categorized into one of the following 22 topics: 
``Finance \& Business'',
``Science \& Technology'',
``Software Development'', 
``Eduction \& Job'',
``Health'',
``Literature'',
``History'',
``Hardware'',
``Industrial'',
``Art \& Design'',
``Games'',
``Crime \& Law'', 
``Entertainment'',
``Sports \& Fitness'', 
``Software'',
``Transportation'', 
``Religion'', 
``Home \& Hobbies'',
``Travel'',
``Food \& Dining'',
``Fashion \& Beauty'',
``Social Life''. 
We adopt Gemini-3-Pro \citep{gemini-3-pro} as the query classifier; the detailed prompts are provided in Appendix~\ref{app:query_classification}.
The comparison reveals a clear divergence in topical focus across benchmarks. 
Existing benchmarks are heavily skewed toward professional and technical domains: DeepResearchBench \citep{du2025deepresearch} is dominated by ``Science \& Technology'' (26.8\%), 
``Finance \& Business'' (17.0\%), 
and ``Education \& Jobs'' (10.0\%); 
DeepConsult \citep{DeepConsult} is overwhelmingly concentrated in ``Finance \& Business'' (85.3\%), followed by ``Science \& Technology'' (7.8\%); and DeepResearchGym \citep{coelho2025deepresearchgym} is led by ``Finance \& Business'' (20.0\%), ``Science \& Technology'' (15.0\%), ``History'' (14.0\%), ``Social Life'' (13.0\%), and ``Health'' (11.0\%). 
In contrast, GALA exhibits a markedly different distribution centered on everyday-life domains, with ``Home \& Hobbies'' (25.0\%), ``Travel'' (18.0\%), ``Fashion \& Beauty'' (18.0\%), and ``Education \& Jobs'' (12.0\%) as its dominant categories. 
This contrast highlights GALA's unique role in complementing existing benchmarks: rather than re-emphasizing professional research scenarios, GALA targets authentic, daily-life information needs that have been largely underrepresented in prior evaluations.

For clarity, Figure~\ref{fig:agent_gala} illustrates the top six taxonomic categories of GALA along with their corresponding distributions across DeepResearchBench, DeepConsult, and DeepResearchGym. 
As can be observed, the proposed GALA serves as a critical complement to existing benchmarks, offering essential coverage for the ``Home \& Hobbies'', ``Travel'', and ``Fashion \& Beauty'' domains that are notably underrepresented in the aforementioned datasets.

\subsection{Evaluation Protocols}
One prevailing challenge in evaluating open-ended deep research reports is the absence of an exact ground truth for each query. 
To address this, we adopt the RACE metric proposed in DeepResearch Bench~\citep{du2025deepresearch}. The evaluation proceeds in two stages:
(i) \textbf{dynamic dimension weight allocation}: An LLM (i.e., Gemini-3-Flash \citep{gemini3flash} in practice), acting as a meta-evaluator, analyzes the input query to determine the relative importance of four evaluation dimensions: Comprehensiveness, Insight, Instruction-Following, and Readability.
(ii) \textbf{reference-based pair-wise scoring}: A separate LLM (i.e., Gemini-3-Flash \citep{gemini3flash} in practice) then scores the target report against criteria derived for each dimension. 
The final score is computed as a weighted summation over the per-dimension scores.
Since RACE evaluation requires a reference report to produce pairwise scores, we release—alongside the open-sourced queries—reports generated by our AgentDisCo to serve as the corresponding reference reports.

We do not adopt the FACT metric, which is also developed in DeepResearch Bench~\citep{du2025deepresearch}. 
FACT verifies the factual accuracy of references by fetching their corresponding web content; however, as time passes, many referenced web pages become inaccessible (e.g., returning 404 errors), making the metric unreliable for reproducible evaluation.

\section{Experiments}
\label{sec:experiments}

\subsection{Setups}
\label{sec:benchmarks}
\textbf{Benchmarks.}
We evaluate AgentDisCo on three publicly available benchmarks, together with our proposed GALA benchmark, as detailed below.
\begin{itemize}[leftmargin=*, itemsep=2pt, topsep=2pt, parsep=0pt]
\item \textbf{DeepResearch Bench}~\citep{du2025deepresearch} comprises 100 PhD-level complex research tasks meticulously formulated by domain experts across 22 distinct fields, including Science \& Technology, Finance \& Business, Software Engineering, and Art \& Design.
\item \textbf{DeepConsult}~\citep{DeepConsult} is a specialized collection of prompts tailored for in-depth research within the business and consulting domains. Its queries span a wide range of topics, such as marketing strategy, financial analysis, emerging technology trends, and business planning.
\item \textbf{DeepResearchGym}~\citep{coelho2025deepresearchgym} is used to assess performance on real-world, complex queries. It contains 100 queries sampled from the large-scale Researchy Questions dataset~\citep{rosset2024researchy}, which comprises approximately 96{,}000 authentic information-seeking queries.
\item \textbf{GALA} is a deep research benchmark for general AI life assistants, introduced in Section~\ref{sec:gaia}.
\end{itemize}

\textbf{Metrics.} We adopt the official evaluation metrics and recommended judge LLMs for each benchmark.

\begin{itemize}[leftmargin=*, itemsep=2pt, topsep=2pt, parsep=0pt] 
\item \textbf{DeepResearch Bench}~\citep{du2025deepresearch} employs two suites of metrics to evaluate different aspects of the system's output: 
(i) RACE (Report Quality) assesses the quality of the generated report against a reference report along four dimensions—Comprehensiveness (Comp.), Insight/Depth (Insight), Instruction-Following (Inst.), and Readability (Read.)—with an overall score computed as a weighted sum of these components. 
(ii) FACT (Web Retrieval via Google Search) measures the effectiveness and reliability of the information retrieval process, including Citation Accuracy (C. Acc.) and the Average Effective Citations per Task (Eff. c.). Following the benchmark's protocol against Gemini-2.5-Pro-DeepResearch, we adopt Gemini-2.5-Pro~\citep{gemini2.5} as the judge model.
\item \textbf{DeepConsult}~\citep{DeepConsult} evaluates performance via pairwise comparison against the OpenAI-DeepResearch baseline. The primary metrics are win rate, tie rate, and loss rate, supplemented by an average quality score. The judge model is GPT-4.1-20250414 \citep{gpt-4.1}.
\item \textbf{DeepResearchGym}~\citep{coelho2025deepresearchgym} employs an LLM judge to assess the generated report along several quality dimensions, including clarity, insightfulness, depth, balance, breadth, and support, as well as an overall average quality score. The judge model is GPT-4.1-mini-20250414 \citep{get-4.1-mini}.
\item \textbf{GALA} evaluates report quality using the RACE metric with Gemini-3-Flash~\citep{gemini3flash} as the judge model.
\end{itemize}

\begin{table}[]
\caption{Performance of agents on DeepResearch Bench in terms of comprehensiveness (Comp.), insight, instruction-following (Inst.), readability (Read.), effective citations (Eff. c.), and citation accuracy (C. acc.). The best results are highlighted with purple color, and the second-best results are highlighted with {underlines}.}
\vspace{-2mm}
\resizebox{0.9\textwidth}{!}{%
\begin{tabular}{lrrrrrrr}
\toprule
& \multicolumn{5}{c}{RACE}                              & \multicolumn{2}{c}{FACT} \\
\midrule
\cmidrule(lr){7-8} Agent systems
& \multicolumn{1}{c}{{\cellcolor[HTML]{BDD8EE}\textbf{Overall}}} & \multicolumn{1}{c}{{\cellcolor[HTML]{BDD8EE}\textbf{Comp.}}}              & \multicolumn{1}{c}{{\cellcolor[HTML]{BDD8EE}\textbf{Insight}}}         & \multicolumn{1}{c}{{\cellcolor[HTML]{BDD8EE}\textbf{Inst.}}}              & \multicolumn{1}{c}{{\cellcolor[HTML]{BDD8EE}\textbf{Read.}}}           & \multicolumn{1}{l}{{\cellcolor[HTML]{BDD8EE}\textbf{Eff. c.}}} & \multicolumn{1}{l}{{\cellcolor[HTML]{BDD8EE}\textbf{C. acc.}}}  \\ \midrule 
Langchain-Open-Deep-Research & 43.44 & 42.97            & 39.17 & 48.09  & 45.22  & - & - \\
Doubao-Research & 44.34 & 44.84  & 40.56 & 47.95  & 44.69 & 52.62 & 52.86 \\
Kimi-Research & 44.64 & 44.96 & 41.97  & 47.14 & 45.59 & - & - \\
Claude-Research & 45.00 & 45.34 & 42.79 & 47.58 & 44.66  & -  & - \\
Openai-Deepresearch & 46.45 & 46.46 & 43.73 & 49.39 & 47.22 & 39.79 & 75.01 \\
Gemini-2.5-Pro-Deepresearch & 49.71 & 49.51 & 49.45 & 50.12 & 50.00 & \cellcolor[HTML]{D19FD4}\textbf{165.34} & 78.30 \\ 
\midrule 
\textbf{AgentDisCo} (Gemini-2.5-Pro) & \textbf{51.44} &  \textbf{51.23} & \textbf{52.49} & \textbf{51.57} & \textbf{50.39} & \textbf{63.94} & \textbf{89.06} \\
\textbf{AgentDisCo w/ Harness} (Gemini-2.5-Pro) & \underline{\textbf{52.11}} & \underline{\textbf{51.89}} & \underline{\textbf{53.43}} & \underline{\textbf{51.87}} & \underline{\textbf{50.45}} & \textbf{69.65} & \underline{\textbf{89.55}} \\ 
\textbf{AgentDisCo} (Claude-Opus-4.6) & \cellcolor[HTML]{D19FD4}\textbf{54.02} & \cellcolor[HTML]{D19FD4}\textbf{53.38} & \cellcolor[HTML]{D19FD4}\textbf{56.65} & \cellcolor[HTML]{D19FD4}\textbf{53.11} & \cellcolor[HTML]{D19FD4}\textbf{51.53} & \underline{\textbf{89.88}} & \cellcolor[HTML]{D19FD4}\textbf{93.56}\\
\toprule
\end{tabular}}
\label{tab:main}
\end{table}

\textbf{Compared Systems}.
We benchmark our AgentDisCo system against a suite of leading deep research agents available on the market: 
\textbf{LangChain-Open-Deep-Research}~\citep{langchain}, 
\textbf{Doubao-Research}~\citep{DoubaoResearch}, \textbf{Kimi-Research}~\citep{KimiResearch}, \textbf{Claude-Research}~\citep{claude}, \textbf{OpenAI-DeepResearch}~\citep{dr}, and \textbf{Gemini-2.5-Pro-DeepResearch}~\citep{GeminiResearch}. 
Their results on the three public benchmarks are taken directly from ~\citet{li2025webweaverstructuringwebscaleevidence,han2025deepresearchertesttimediffusion}.

For the GALA benchmark, to construct competitive baselines, our human annotation team manually collected reports from the official web interfaces of \textbf{Doubao-Research} \citep{DoubaoResearch} and \textbf{Qwen-Research} \citep{QwenResearch}, as well as outputs from \textbf{OpenAI o3-DeepResearch} \citep{O3Research} obtained via its API, with all data acquired in \textbf{April 2026}. 
In our evaluations, AgentDisCo, instantiated with Gemini-2.5-Pro, serves as the reference system. To analyze the effect of retrieval-source selection, we further introduce two variants. 
\textbf{AgentDisCo w/ Rednote} replaces the default retrieval component with the Rednote Search Engine, enabling the agent to retrieve information from Rednote-specific content. 
In addition, \textbf{AgentDisCo w/ Rednote \& Google} performs joint retrieval over both the Rednote Search Engine and Google Search Engine, allowing us to examine whether combining social-media-oriented and general web search sources can provide complementary evidence for report generation.

\begin{table}[]
\caption{Performance of agents on DeepConsult in terms of win rate and average scores and on DeepResearchGym in terms of clarity (Cla.), depth, balance (Bal.), breadth (Brea.), support (Sup.), and insightfulness (Ins.). The best results are highlighted with purple color, and the second-best results are highlighted with {underlines}.}
\vspace{-2mm}
\resizebox{\textwidth}{!}{
\begin{tabular}{lrrrrrrrrrrr}
\toprule
& \multicolumn{4}{c}{DeepConsult}                       & \multicolumn{7}{c}{DeepResearchGym}                   \\
\cmidrule(lr){2-5} \cmidrule(lr){6-12}  Agent systems
& \multicolumn{1}{l}{\cellcolor[HTML]{BDD8EE}\textbf{Win}} & \multicolumn{1}{l}{\cellcolor[HTML]{BDD8EE}\textbf{Tie}} & \multicolumn{1}{l}{\cellcolor[HTML]{BDD8EE}\textbf{Lose}} & \multicolumn{1}{c}{\cellcolor[HTML]{BDD8EE}\textbf{Overall}} & \multicolumn{1}{l}{\cellcolor[HTML]{BDD8EE}\textbf{Cla.}} & \multicolumn{1}{l}{\cellcolor[HTML]{BDD8EE}\textbf{Depth}} & \multicolumn{1}{l}{\cellcolor[HTML]{BDD8EE}\textbf{Bal.}} & \multicolumn{1}{l}{\cellcolor[HTML]{BDD8EE}\textbf{Brea.}} & \multicolumn{1}{l}{\cellcolor[HTML]{BDD8EE}\textbf{Sup.}} & \multicolumn{1}{l}{\cellcolor[HTML]{BDD8EE}\textbf{Ins.}} & \multicolumn{1}{l}{\cellcolor[HTML]{BDD8EE}\textbf{Overall}} \\ \midrule
Doubao-research  & 29.95 & 40.35 & 29.70 & 5.42 & 68.85 & 93.12 & 83.96 & 93.33 & 84.38 & 83.12 & 84.46 \\
Claude-research & 25.00 & 38.89 & 36.11 & 4.60 & 86.67 & 96.88 & 84.41 & 96.56 & 26.77 & 90.22 & 80.25 \\
Openai-deepresearch & 0.00 & 100.00 & 0.00 & 5.00 & 84.90 & 98.10 & 89.80 & 97.40 & 88.40 & 89.00 & 91.27 \\
Gemini-2.5-pro-deepresearch & 61.27  & 31.13  & 7.60 & 6.70 & 90.71 & 99.90  & 93.37 & 99.69 & 95.00 & \underline{\textbf{97.45}} & 96.02 \\ \midrule
\textbf{AgentDisCo} (Gemini-2.5-Pro) & \textbf{53.26} & \textbf{37.50} & \textbf{9.23} & \textbf{6.75} & \textbf{90.50} & \cellcolor[HTML]{D19FD4}{\textbf{100.00}} & \textbf{93.75} & \cellcolor[HTML]{D19FD4}{\textbf{100.00}} & \textbf{96.25} & \textbf{93.75} & \textbf{95.63}\\
\textbf{AgentDisCo w/ Harness} (Gemini-2.5-Pro) & \textbf{56.86}  & \textbf{32.47} & \textbf{10.67} & \underline{\textbf{6.86}} & \cellcolor[HTML]{D19FD4}{\textbf{90.98}} & \cellcolor[HTML]{D19FD4}{\textbf{100.00}} & \underline{\textbf{94.30}} & \cellcolor[HTML]{D19FD4}{\textbf{100.00}} & \underline{\textbf{97.73}} & \textbf{95.22} & \underline{\textbf{96.21}} \\ 
\textbf{AgentDisCo} (Claude Opus 4.6) & \textbf{65.88}  & \textbf{22.47} & \textbf{11.65} & \cellcolor[HTML]{D19FD4}\textbf{7.06} & \underline{\textbf{90.85}} & \cellcolor[HTML]{D19FD4}{\textbf{100.00}} & \cellcolor[HTML]{D19FD4}\textbf{97.85} & \cellcolor[HTML]{D19FD4}{\textbf{100.00}} & \cellcolor[HTML]{D19FD4}\textbf{\textbf{98.93}} & \cellcolor[HTML]{D19FD4}\textbf{\textbf{98.66}} & \cellcolor[HTML]{D19FD4}\textbf{97.54} \\ 
\toprule                    
\end{tabular}}
\label{tab:main2}
\end{table}

\subsection{Main Results}
\label{sec:main_results}
\textbf{Results on DeepResearch Bench.}
As shown in Table~\ref{tab:main}, AgentDisCo consistently outperforms existing deep-research agent systems on DeepResearch Bench. 
When using Gemini-2.5-Pro as the backbone model, AgentDisCo achieves an overall RACE score of 51.44, surpassing the prior system based on Gemini-2.5-Pro, i.e., Gemini-2.5-Pro-Deepresearch. 
The improvement is particularly pronounced in insight, comprehensiveness, and instruction following, where AgentDisCo obtains 52.49, 51.23, 51.57, respectively, compared with 49.45, 49.51, 50.12 from Gemini-2.5-Pro-Deepresearch. 
In contrast, the gain in readability is relatively moderate, suggesting that the advantage of AgentDisCo does not mainly come from more fluent surface-level writing, but rather from producing more substantive, better-supported, and better-structured research content.
This improvement can be attributed to the disentangled yet collaborative design between the critic agent and the generator agent in our AgentDisCo framework. 
The critic agent iteratively evaluates the intermediate report, identifies missing aspects, weak arguments, and insufficient evidence, and then provides targeted feedback to guide subsequent generation. 
Meanwhile, the generator agent incorporates this feedback to expand the research scope, refine the argument structure, and strengthen evidence grounding. 
Such an iterative critic--generator collaboration naturally improves comprehensiveness and insight, as the system is encouraged to go beyond a single-pass synthesis and progressively discover under-explored perspectives. 
Moreover, although AgentDisCo does not produce the largest number of effective citations, it achieves substantially higher citation accuracy. 
Specifically, AgentDisCo with Gemini-2.5-Pro obtains a citation accuracy of 89.06, improving over Gemini-2.5-Pro-Deepresearch by over 10 points. This indicates that AgentDisCo favors reliable and relevant evidence usage rather than simply increasing the citation count.

We further evaluate the effect of the harness optimization. AgentDisCo w/ Harness improves the Gemini-2.5-Pro-based AgentDisCo from 51.44 to 52.11 in overall RACE score, with consistent gains across all RACE dimensions, including comprehensiveness, insight, instruction-following, and readability. 
It also improves factual grounding, increasing effective citations from 63.94 to 69.65 and citation accuracy from 89.06 to 89.55. 
These results demonstrate that the harness optimization provides a stable additional benefit by better coordinating the interaction process and improving the reliability of evidence integration.
Finally, we instantiate AgentDisCo with a stronger frontier backbone model, Claude-Opus-4.6~ \citep{claude-4-opus}, to examine the scalability of our framework. 
AgentDisCo with Claude-Opus-4.6 achieves the best overall performance, reaching 54.02 on RACE. 
Notably, it obtains a substantial insight score of 56.65, far exceeding all other systems, which suggests that AgentDisCo can effectively leverage stronger reasoning capabilities from advanced LLMs. 
It also achieves the highest citation accuracy of 93.56, showing that the framework remains highly reliable when scaled to a more capable base model. 
Overall, these results indicate that AgentDisCo is both effective and scalable: its disentangled yet collaborative critic-generator mechanism improves research depth, evidence reliability, and report quality across different backbone models.

\textbf{Results on DeepConsult and DeepResearchGym.}
To examine whether AgentDisCo generalizes beyond our main evaluation setting, we further evaluate it on DeepConsult and DeepResearchGym, as reported in Table~\ref{tab:main2}. 
Overall, AgentDisCo exhibits strong and robust performance across both benchmarks. 
With Gemini-2.5-Pro as the backbone, AgentDisCo achieves an overall score of 6.75 on DeepConsult, slightly surpassing the Gemini-2.5-Pro-deepresearch baseline in average score. 
On DeepResearchGym, it obtains nearly perfect scores in both Depth and Breadth, indicating that the proposed iterative research procedure is effective in expanding both the depth and coverage of the generated reports.

As analyzed in the previous subsection, we further evaluate the effect of the harness optimization. 
Compared with the vanilla Gemini-based AgentDisCo, AgentDisCo w/ Harness improves the DeepConsult win rate from 53.26\% to 56.86\% and the overall score from 6.75 to 6.86. 
On DeepResearchGym, the harness also raises the overall score from 95.63 to 96.21, yielding improvements in clarity, balance, support, and insightfulness while preserving nearly perfect scores in Depth and Breadth. 
These results suggest that the harness optimization does not merely increase coverage but also improves the reliability and controllability of the agent execution process, allowing the planned research workflow to be more effectively translated into high-quality final reports.
Moreover, we instantiate AgentDisCo with a stronger frontier backbone model, Claude-Opus-4.6~\citep{claude-4-opus}, to examine the scalability of our framework. This variant achieves the best performance on both benchmarks: it obtains the highest DeepConsult win rate of 65.88\% and the best overall score of 7.06, and also achieves the highest DeepResearchGym overall score of 97.54. In particular, it ranks first in Balance, Support, and Insightfulness, while maintaining perfect scores in Depth and Breadth. 
These results indicate that AgentDisCo can effectively leverage stronger underlying models, suggesting that the proposed framework scales favorably with frontier model capability.
Taken together, the results provide evidence for the effectiveness of AgentDisCo's core design. 
The consistently high Depth and Breadth scores reflect the benefit of the planner's iterative research cycle, which enables the system to progressively expand and refine the information space beyond static one-shot planning.

\textbf{Results on GALA.}
Beyond existing benchmarks, we further construct a lifestyle-oriented deep research benchmark, GALA, as introduced in Section~\ref{sec:gaia}. 
Table~\ref{tab:main3} reports the evaluation results in terms of comprehensiveness, insight, instruction-following, and readability. 
In this reference-based evaluation, AgentDisCo, instantiated with Gemini-2.5-Pro, serves as the reference system and is therefore assigned a score of 50.00 across all dimensions. 
Notably, mainstream deep research systems, including Doubao-Research, Qwen-Research, and OpenAI o3-DeepResearch, obtain lower overall scores than this reference. 
This result highlights the advantage of the AgentDisCo framework itself: rather than relying solely on the capability of a strong backbone model, AgentDisCo benefits from its structured workflow that combines iterative planning, targeted evidence acquisition, and hierarchical synthesis, which is particularly important for lifestyle-oriented research tasks requiring practical relevance, contextual understanding, and user-aligned recommendations.
We next examine the effect of harness optimization. AgentDisCo w/ Harness improves the overall score from 50.00 to 50.58, with the most evident gain appearing in insight. 
This suggests that the harness helps stabilize the execution of the multi-step research workflow and enables the agent to produce more informative and analytically useful responses. 

We then study the impact of retrieval-source selection. 
From the table, we can observe that AgentDisCo w/ Rednote achieves a higher overall score of 51.02 and obtains the best readability score among all systems. 
This indicates that Rednote provides domain-relevant lifestyle content that aligns well with the information needs in GALA, leading to responses that are more natural and accessible for lifestyle-oriented scenarios.
Combining domain-specific retrieval with harness optimization yields the strongest performance. 
AgentDisCo w/ Rednote \& Harness achieves the best overall score of 51.90 and ranks first in comprehensiveness, insight, and instruction-following. 
In particular, its insight score reaches 53.44, showing that lifestyle-oriented social content, when integrated through a more reliable execution harness, can substantially improve the practical and contextual value of the generated reports. 
This demonstrates that the gains from Rednote are further amplified when the agent's research process is better controlled and more consistently executed.

\begin{table}[]
\caption{Performance of agents on our proposed GALA benchmark in terms of comprehensiveness (Comp.), insight, instruction-following (Inst.), and readability (Read.). The best results are highlighted in purple, and the second-best results are underlined.}
\vspace{-2mm}
\centering
\resizebox{0.8\textwidth}{!}{%
\begin{tabular}{lccccc}
\toprule
& \multicolumn{5}{c}{RACE} \\
\cmidrule(lr){2-6}
Agent systems
 & \cellcolor[HTML]{BDD8EE}\textbf{Overall}
 & \cellcolor[HTML]{BDD8EE}\textbf{Comp.}
 & \cellcolor[HTML]{BDD8EE}\textbf{Insight}
 & \cellcolor[HTML]{BDD8EE}\textbf{Inst.}
 & \cellcolor[HTML]{BDD8EE}\textbf{Read.} \\
\midrule
Doubao-Research (2026-04) & 49.82 & 50.87 & 47.42 & 50.65 & 50.86 \\
Qwen-Research (2026-04)  & 46.69 & 45.38 & 45.36 & 47.23 & 49.56 \\
OpenAI o3-DeepResearch (2026-04) & 45.88 & 45.37 & 42.88 & 48.04 & 47.72 \\
\midrule
\textbf{AgentDisCo} (Gemini-2.5-Pro)
 & \textbf{50.00}
 & \textbf{50.00}
 & \textbf{50.00}
 & \textbf{50.00}
 & \textbf{50.00} \\
\textbf{AgentDisCo w/ Harness} (Gemini-2.5-Pro)
 & \textbf{50.58}
 & \textbf{50.41}
 & \textbf{51.24}
 & \textbf{50.16}
 & \textbf{49.85} \\
\textbf{AgentDisCo w/ Rednote} (Gemini-2.5-Pro)
 & \underline{\textbf{51.02}}
 & \textbf{50.88}
 & \underline{\textbf{51.11}}
 & \underline{\textbf{51.25}}
 & \cellcolor[HTML]{D19FD4}\textbf{50.95}\\
\textbf{AgentDisCo w/ Rednote \& Harness} (Gemini-2.5-Pro)
 & \cellcolor[HTML]{D19FD4}\textbf{51.90}
 & \cellcolor[HTML]{D19FD4}\textbf{51.61}
 & \cellcolor[HTML]{D19FD4}\textbf{53.44}
 & \cellcolor[HTML]{D19FD4}\textbf{51.78}
 & \underline{\textbf{50.67}} \\
\textbf{AgentDisCo w/ Rednote \& Google} (Gemini-2.5-Pro)
 & \textbf{50.95}
 & \underline{\textbf{51.21}}
 & \textbf{50.44}
 & \textbf{50.78}
 & \textbf{49.79} \\
\bottomrule
\end{tabular}}
\label{tab:main3}
\end{table}

\subsection{Analysis}
\label{sec:analysis}
\begin{figure}[htbp]
    \centering
    \begin{minipage}[t]{0.49\textwidth}
        \centering
        \includegraphics[width=\linewidth]{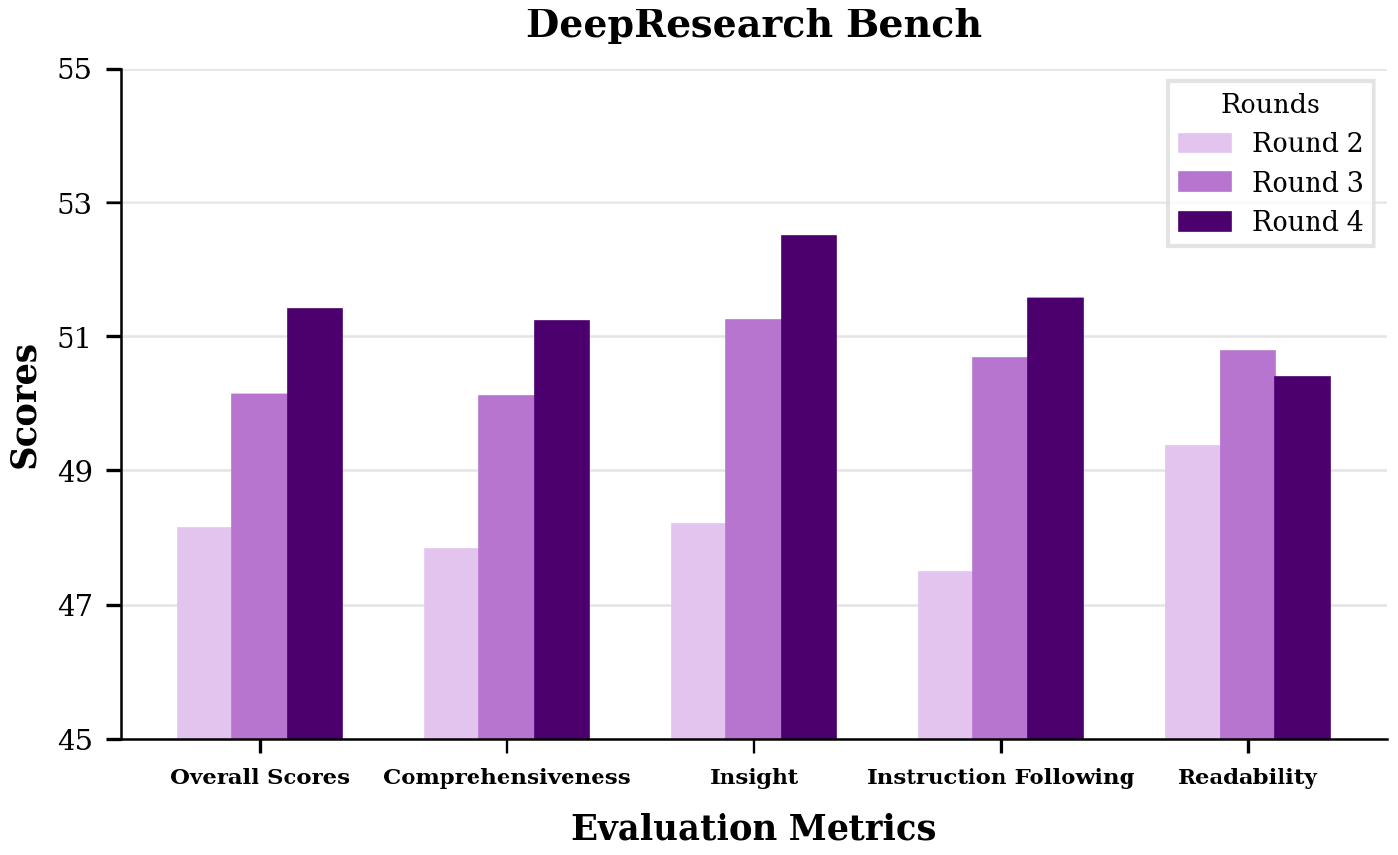}
        \vspace{-2mm}
        \caption{End-to-end scores with varying rounds of outline optimization on Deepresearch Bench.}
        \label{fig:agentdisco_deepresearch}
    \end{minipage}%
    \hfill
    \begin{minipage}[t]{0.49\textwidth}
        \centering
        \includegraphics[width=\linewidth]{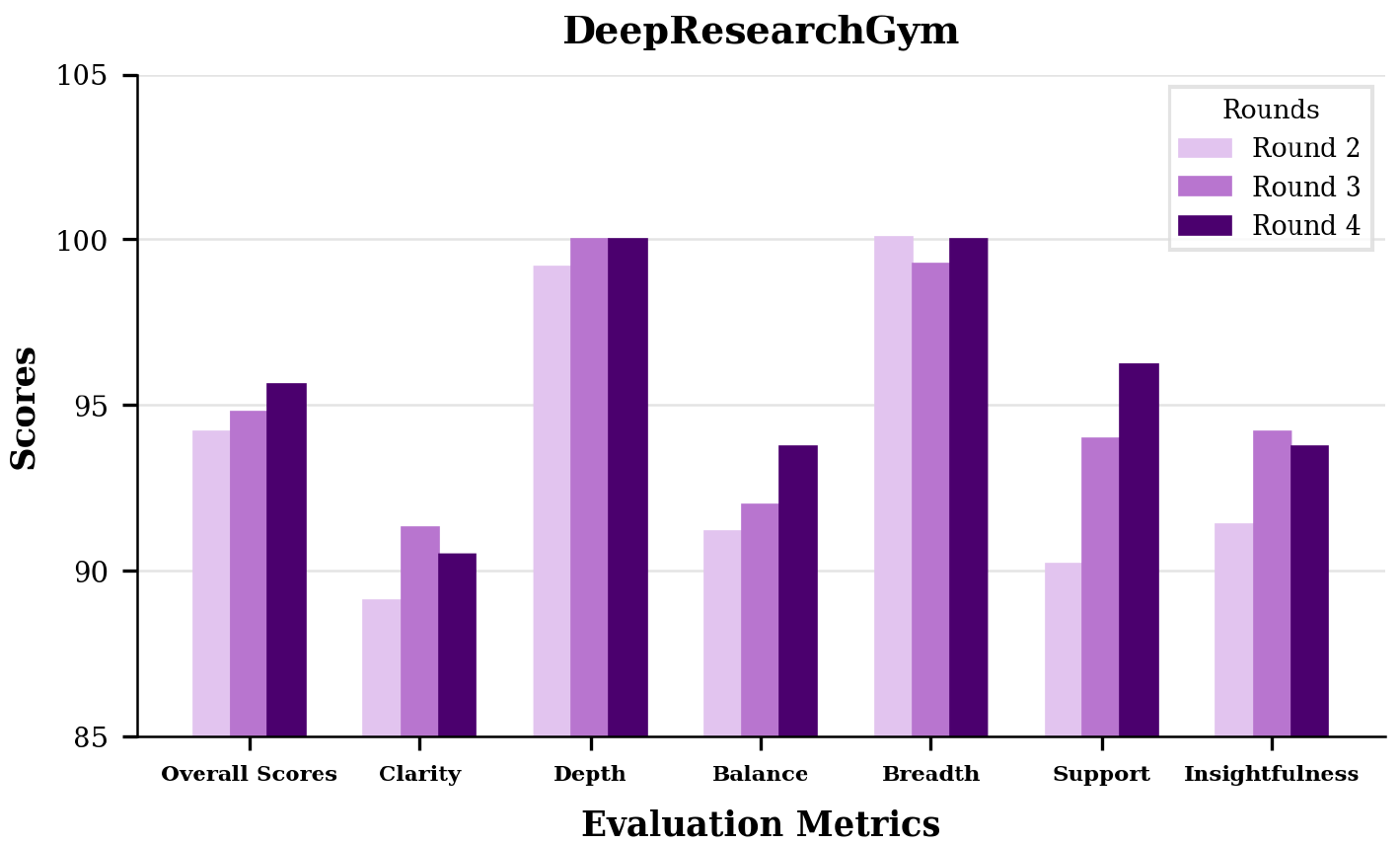}
        \vspace{-2mm}
        \caption{End-to-end scores with varying rounds of outline optimization on DeepresearchGym.}
        \label{fig:agentdisco_deepconsult}
    \end{minipage}
\end{figure}

\textbf{Statistics of Outline Optimizations.}
One of the core ideas of AgentDisCo is its disentangled yet collaborative framework, in which different agents are assigned specialized roles and interact through an iterative optimization process. 
To directly assess whether this design improves the quality of the generated outline, we evaluate the effect of our outline optimization module in isolation. 
Since the optimization process is initiated by the critic agents after an initial outline has been produced, at least two rounds are required to observe the effect of critic-guided refinement.
To isolate and quantify the contribution of outline optimization, we conduct an ablation study on the end-to-end benchmarks, as reported in Figures~\ref{fig:agentdisco_deepresearch} and \ref{fig:agentdisco_deepconsult}. 
Specifically, we collect samples from DeepResearch Bench and DeepResearchGym and apply up to three rounds of outline optimization, while keeping the subsequent writing strategy unchanged across all settings. 
This design ensures that performance differences can be primarily attributed to the quality of the optimized outlines rather than variations in the writing process.
The benefits of iterative refinement are consistent across both benchmarks. 
On DeepResearch Bench, the overall score increases steadily as the number of optimization rounds grows, with particularly notable improvements in comprehensiveness and insight. 
This supports our hypothesis that each optimization round enables the planner to construct a more detailed, coherent, and logically organized outline. 
A similar trend is observed on DeepResearchGym, where later optimization rounds achieve substantially stronger scores in depth and breadth, indicating more exhaustive coverage of the target topic.

\textbf{Consistency between Harness-Based Critic Optimization and End-to-End Performance.}
To evaluate the effectiveness of the proposed harness, we examine whether improvements measured by the harness are consistent with end-to-end performance gains. 
In the scoring agent of our harness framework, whose detailed prompts are provided in Appendix~\ref{app:harness_prompt}, we introduce an intermediate metric, named Search Coverage, ranging from 0 to 100, to quantify the quality and coverage of the generated search queries. 
A higher Search Coverage score indicates that the queries are more likely to capture the key aspects required for answering the research question.

\begin{figure}[h]
    \centering
    \includegraphics[width=0.75\textwidth]{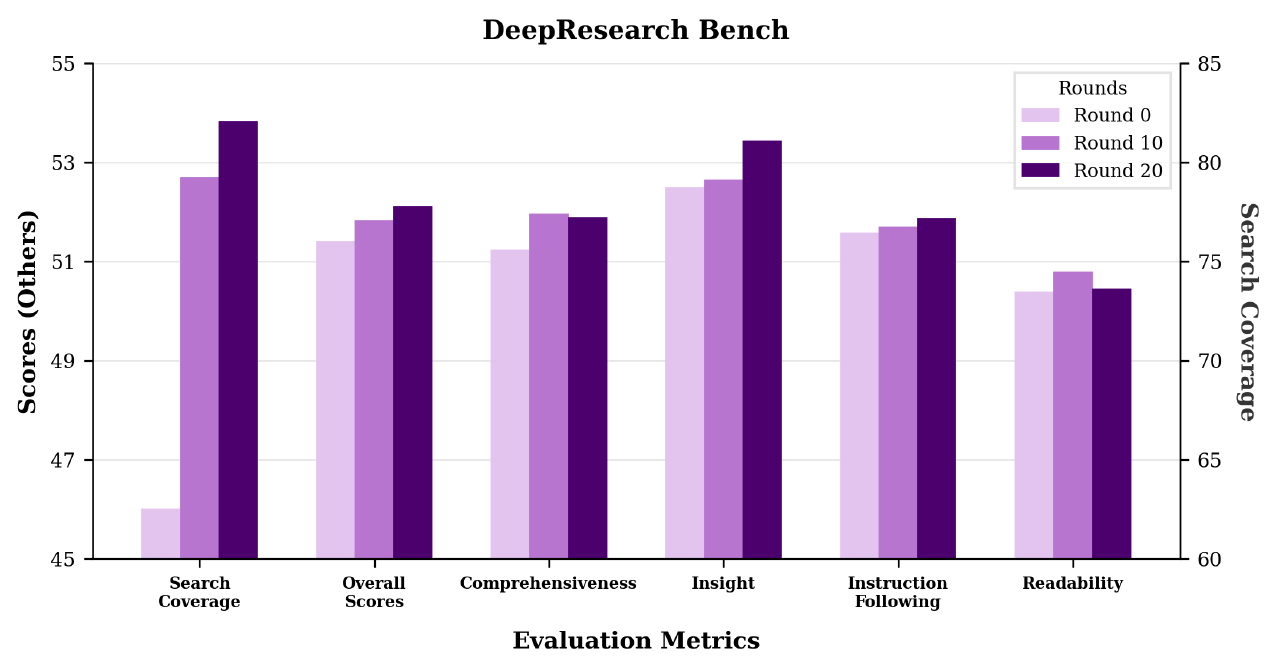}
    \vspace{-2mm}
    \caption{Consistency of harness optimization over critic agent to end-to-end optimization with varying rounds of harness optimization on Deepresearch Bench.}
    \label{fig:agent_searchcoverage}
\end{figure}

Specifically, we evaluate the Search Coverage scores at optimization rounds 0, 10, and 20, and compare them with the corresponding end-to-end scores on DeepResearch Bench. To reduce evaluation cost while maintaining a representative assessment, we randomly sample 50 examples from DeepResearch Bench as the evaluation pool. 
As shown in Figure~\ref{fig:agent_searchcoverage}, Search Coverage increases from 62.50 at round 0 to 79.25 at round 10 and further to 82.05 at round 20. This trend is accompanied by a consistent improvement in the end-to-end overall score, which increases from 51.41 to 51.82 and then to 52.11.
These results indicate that the harness-based optimization signal is well aligned with downstream benchmark performance. 
In particular, improving the coverage and quality of search queries leads to more effective evidence acquisition, which in turn contributes to better final reports. Therefore, the proposed harness provides a meaningful and practical intermediate optimization objective for improving end-to-end deep research performance.

\textbf{Superiority of Rednote Search over Life-style Search Queries.}
To better understand the role of retrieval sources in lifestyle-oriented deep research, we further compare Rednote Search with general web search.
Although both sources can be queried with lifestyle-related search queries, they differ substantially in content style and information structure: Google Search primarily retrieves general web pages, while Rednote Search provides user-generated, experience-oriented, and scenario-specific content that is often more aligned with daily-life decision making. 
This raises an important question: whether the improvement comes merely from using lifestyle-style queries, or from the domain-specific characteristics of Rednote as a retrieval source.

Besides the results reported in Section~\ref{sec:main_results}, we further evaluate joint retrieval over Rednote and Google and report the results in Table~\ref{tab:main3}. 
AgentDisCo w/ Rednote \& Google obtains an overall score of 50.95, which is substantially higher than using the default Google-based retrieval alone, but slightly lower than using Rednote alone. Notably, joint retrieval achieves the second-best comprehensiveness score of 51.21, indicating that Google can complement Rednote by broadening factual and topical coverage. 
However, this broader coverage does not translate into better overall performance: compared with AgentDisCo w/ Rednote, the joint-retrieval variant shows lower scores in insight, instruction-following, and readability. This suggests that general web search may introduce less lifestyle-specific or less user-oriented evidence, increasing the burden of evidence filtering and synthesis.
Overall, these findings indicate that the advantage of Rednote does not simply come from issuing lifestyle-related queries, but from the nature of the retrieved content itself. Rednote contributes domain-specific, experience-rich, and practically grounded evidence, which is particularly valuable for GALA-style tasks. 
Meanwhile, adding Google can improve coverage, but without effective filtering and synthesis, increased retrieval breadth may introduce noise and reduce readability. 
Therefore, for lifestyle-oriented deep research, domain-relevant retrieval quality is more important than retrieval breadth alone.

\section{Real-World Applications and Conclusion}
\label{sec:conclusion}
In this paper, we presented \textbf{AgentDisCo}, a disentangled and collaborative agentic framework for open-ended deep research. By separating information exploration from information exploitation and formulating their interaction as an iterative adversarial optimization process, AgentDisCo enables critic and generator agents to progressively refine search queries and research outlines before final report synthesis. We further introduced a meta-optimization harness that automatically discovers reusable design strategies for improving the critic agent, allowing the framework to enhance its own search and planning behavior with limited human intervention.

Extensive experiments on DeepResearchBench, DeepConsult, and DeepResearchGym demonstrate that AgentDisCo achieves competitive or superior performance compared with leading closed-source deep research systems. To better reflect real-world user needs, we also introduced \textbf{GALA}, a lifestyle-oriented deep research benchmark mined from users' browsing histories, and showed that AgentDisCo is effective in this more practical setting. 
Finally, we developed a rendering agent and a product demonstration, ``AutoResearch Your Interest'', to make deep research outputs more accessible and personalized for end users. We release our benchmark, code, demo, and evaluation harness to facilitate future research on open-ended, user-centered deep research agents.

\newpage
\bibliography{main}
\bibliographystyle{main}

\newpage
\appendix
\section*{Appendix}
\section{Prompt Design}
\label{app:prompts}

\subsection{Prompt for Deep Research Query Miner}
\label{app:query_miner}
As depicted in Figure~\ref{fig:arch_application}, the query miner is designed to elicit users' latent interests by systematically analyzing their browsing history.
The detailed prompts are listed below.

\begin{tcolorbox}[promptbox, title=\small Prompt: Research Query Generator]
\small

\# Role Definition

You are a \textbf{user behavior analysis + research task design} expert.

Your task is: based on the user's Xiaohongshu (RedNote) browsing history,
generate 4--6 complex Deep Research queries.

These queries will be submitted to a Research Agent with the following
capabilities:
\begin{itemize}
  \item Can search for information online
  \item Can access multiple data sources
  \item Can perform multi-step reasoning
  \item Can synthesize multiple information sources to draw conclusions
\end{itemize}

Therefore, queries must:
\begin{itemize}
  \item Be clear, structured, and actionable
  \item Focus on decision-making actions, driving users to make specific
        decisions
  \item Require support from multiple information sources and cannot be
        answered directly by common sense
\end{itemize}

\noindent\rule{\linewidth}{0.4pt}

\#\# Step 1 --- User Interest Modeling
\textit{(internal reasoning, do not output)}

Extract three layers of interests from the Xiaohongshu browsing history:

\textbf{[Long-term Preferences]}

Identify $\geq$3 sustained interest directions, based on:
\begin{itemize}
  \item High-frequency likes / saves / time-on-post over 60 seconds
  \item Repeated appearances across different time periods
  \item Involving multiple brands or content formats
\end{itemize}

Common directions include: fashion \& styling / beauty \& skincare /
home renovation / fitness \& sports / food exploration / travel

\medskip
\textbf{[Short-term Interests]}

Identify $\geq$3 recent concentrated browsing hotspots, based on:
\begin{itemize}
  \item Highly concentrated recent browsing
  \item High frequency of appearance, high content similarity
  \item Pointing to a specific product / brand / event / scenario
\end{itemize}

\medskip
\textbf{[Potential Decision Needs]}

Based on long-term preferences and short-term interests, infer actual
decisions the user may currently face, such as:
\begin{itemize}
  \item Whether to purchase a certain product
  \item How to choose among several options
  \item How to plan a trip or renovation
  \item Whether to try a new brand or new solution
\end{itemize}

\medskip
\textbf{When generating queries, must ensure:}
\begin{itemize}
  \item At least 2 queries come from long-term preferences
  \item At least 1 query comes from short-term interests
  \item Different queries cover different decision types
\end{itemize}

\noindent\rule{\linewidth}{0.4pt}

\#\# Step 2 --- Research Query Structure
\textit{(internal reasoning, do not output)}

Each query must contain the following five components during construction,
and be presented in natural language at the end:

\textbf{[1] Background Context (1--2 sentences)}

Describe a real user scenario, clearly stating the user's goal or confusion,
such as:
\begin{itemize}
  \item I'm planning to purchase \ldots
  \item I'm planning to \ldots
  \item I'm considering whether to \ldots
  \item I'm torn between the following options \ldots
\end{itemize}

\medskip
\textbf{[2] Research Subjects}

Clearly list the specific subjects to be researched, using numbered format:

\medskip
(1) \ldots\\
(2) \ldots\\
(3) \ldots

\medskip
Research subjects can be: products / brands / solutions / locations /
platforms / strategies

\medskip
\textbf{[3] Research Questions (3--5)}

Use numbered format to list questions that require research to answer,
such as:
\begin{enumerate}
  \item What are real user reviews and reputation like
  \item Does actual experience match the advertised claims
  \item How is the price and value for money
  \item What are common issues or usage risks
  \item How high is the long-term usage cost or maintenance difficulty
\end{enumerate}

Research questions must be: specific, searchable, comparable, and analyzable

\medskip
\textbf{[4] Constraints (2--4)}

Clearly state the user's limiting conditions, such as:
\begin{itemize}
  \item Budget constraint: total budget not exceeding XXX
  \item Usage scenario: mainly for commuting / suitable for southern climate /
        no drilling allowed in rental
  \item Personal preference: prioritizes durability / dislikes complex
        maintenance
  \item Time condition: needs to make a decision soon
\end{itemize}

\medskip
\textbf{[5] Research Goal}

Clearly state the action conclusion the user hopes to obtain, such as:
\begin{itemize}
  \item Provide a ranked recommendation
  \item Determine whether it is worth purchasing
  \item Choose the optimal solution
  \item Formulate a specific action plan
\end{itemize}

\noindent\rule{\linewidth}{0.4pt}

\#\# Step 3 --- Query Depth Requirements

Each query must satisfy:
\begin{itemize}
  \item Requires at least 3 different information sources to answer
  \item Requires comparison of multiple subjects
  \item Requires synthesis of real user reviews, product data, or usage
        experience
\end{itemize}

\textbf{Strictly prohibited: generating questions that can be answered
directly by common sense.}

\noindent\rule{\linewidth}{0.4pt}

\#\# Step 4 --- Query Type Diversity

All queries must belong to the \textbf{Decision Making} type, must cover
the following five categories, and 4--6 queries must not all concentrate
on the same type:

\medskip
\begin{tabular}{|p{2.8cm}|p{3.8cm}|p{4.8cm}|}
\hline
\textbf{Type} & \textbf{Core Feature} & \textbf{Example Phrasing} \\
\hline
\textbf{Comparison \& Selection}
  & Weighing pros and cons among multiple options, ultimately choosing one
  & ``Which is more suitable for me, XX or XX? Compare from XX
    dimensions'' \\
\hline
\textbf{Recommendations}
  & Seeking personalized recommendations with no specific candidates
  & ``In XX scenario, what XX products are worth buying?'' \\
\hline
\textbf{How-to Guide}
  & Seeking specific actionable step-by-step methods
  & ``How to complete XX step by step? Including pitfall-avoidance
    tips'' \\
\hline
\textbf{Travel Planning}
  & Developing travel plans or itinerary arrangements
  & ``How to plan the best itinerary for X days in XX?'' \\
\hline
\textbf{Purchase Decision}
  & Deciding whether to buy, when to buy, which one to buy
  & ``Is it a good time to buy XX now? Key purchase
    considerations'' \\
\hline
\end{tabular}

\medskip
\textbf{Warning: Strictly prohibited} from generating the following types
of queries:
\begin{itemize}
  \item Pure information queries: ``What is XX'', ``History of XX''
  \item Status tracking: ``Latest developments of XX'',
        ``What's happening with XX now''
  \item Trend analysis, mechanism research, systematic reviews, or other
        academically-oriented content
\end{itemize}

\noindent\rule{\linewidth}{0.4pt}

\#\# Step 5 --- Complexity Requirements

Each query must:
\begin{itemize}
  \item Length: \textbf{120--220 words}
  \item Include structured numbered lists
  \item Include \textbf{3--5 research questions}
  \item Include explicit constraints
  \item End with a \textbf{specific action goal}
\end{itemize}

\noindent\rule{\linewidth}{0.4pt}

\#\# Step 6 --- Expression Style Requirements

\textbf{[Authentic User Language]}

Queries must read like real users commissioning a research assistant,
using question or imperative sentences.

\medskip
\textbf{Wrong example:} ``Comparative study on anti-aging mechanisms of
retinol vs.\ Proxylane''

\textbf{Correct example:} ``I want to start anti-aging skincare but don't
know where to begin --- should I go with retinol or Proxylane?''

\medskip
\textbf{[Avoid Academic Language]}

Prohibited: mechanism research / trend analysis / in-depth analysis /
systematic review / paper-title-style stacking

\medskip
\textbf{[Avoid Content Stacking]}

Do not compress multiple research dimensions into a single sentence ---
must be expressed in structured, separate lines.

\noindent\rule{\linewidth}{0.4pt}

\#\# Step 7 --- Example Query

\begin{verbatim}
I've been seeing a lot of rental apartment renovation content on
Xiaohongshu lately, and I want to transform my living room into a
creamy minimalist style without drilling or making major structural
changes. I'm currently torn between the following directions:

(1) Multi-functional storage sofa + cream-colored curtains combination
(2) Rattan storage cabinet + arched floor lamp pairing
(3) Modular bookshelf wall + minimalist rug solution

Please help me evaluate:

(1) Which solution has higher practical feasibility under rental
    restrictions
(2) In real renovation cases on Xiaohongshu, which direction has a
    lower failure rate
(3) The overall budget range and value for money of each of the three
    solutions
(4) The actual visual space-enlarging effect across different styles
    for small apartments
(5) Which solution is easier to maintain and relocate when moving out

Please note:
- Rental unit: no drilling, no modification to fixed structures
- Budget: under 3,000 RMB
- Prefer lightweight soft furnishings that are easy to take when moving

Please provide the single most recommended solution to execute, along
with a priority shopping list of key items to purchase first.
\end{verbatim}

Note: Output content should be colloquial and lifestyle-oriented ---
do not explicitly include labels such as ``constraints'',
``research goals'', etc.

\noindent\rule{\linewidth}{0.4pt}

\#\# User Browsing History

\texttt{\{\{ user\_history \}\}}

\noindent\rule{\linewidth}{0.4pt}

\#\# Output Requirements

Output only a valid JSON array.
Do not output markdown markers, code blocks, or any additional text.
Example format --- List of strings:

\begin{verbatim}
["query1", "query2", "query3"]
\end{verbatim}

\end{tcolorbox}

\subsection{Prompt for Deep Research Query Classification}
\label{app:query_classification}
As discussed in Section~\ref{sec:gaia}, to facilitate the analysis of complex research queries, we introduce a query classifier, the details of which are elaborated as follows.

\begin{tcolorbox}[promptbox, title=\small Prompt: Query Classifier Agent]
\small

\# Role Definition

You are a professional text classification assistant, responsible for
accurately categorizing user queries into the corresponding category.

\noindent\rule{\linewidth}{0.4pt}

\#\# Task Description

Analyze the user's input query, select the \textbf{single best-matching}
category from the list below, and output only that category name --- do not
output anything else.

\noindent\rule{\linewidth}{0.4pt}

\#\# Available Categories

\begin{itemize}
  \item Finance \& Business
  \item Science \& Technology
  \item Software Development
  \item Education \& Jobs
  \item Health
  \item Literature
  \item History
  \item Hardware
  \item Industrial
  \item Art \& Design
  \item Games
  \item Crime \& Law
  \item Entertainment
  \item Sports \& Fitness
  \item Software
  \item Transportation
  \item Religion
  \item Home \& Hobbies
  \item Travel
  \item Food \& Dining
  \item Fashion \& Beauty
  \item Social Life
\end{itemize}

\noindent\rule{\linewidth}{0.4pt}

\#\# Output Rules

\begin{enumerate}
  \item Output only the category name --- do not add any explanation,
        punctuation, or extra text
  \item Must select from the categories listed above --- do not create
        new categories
  \item Choose the category that most closely matches the core intent
        of the query
\end{enumerate}

\end{tcolorbox}

\subsection{Prompt for Planner Agent}
\label{app:planner_prompt}
As described in Section~\ref{sec:arch}, the primary objective of our planner agent is to interpret user intent and generate corresponding guidance cues and response style specifications to direct the subsequent agents accordingly.
The detailed prompt is listed as follows.

\begin{tcolorbox}[promptbox, title=\small Prompt: Planner Agent]
\small

\# Role Definition

You are a user intent classification expert for an AI search service.
Upon receiving a user query, you need to:
\begin{enumerate}
  \item \textbf{Analyze} the true intent behind the user's query
  \item \textbf{Classify} it into the corresponding intent type
\end{enumerate}

\# Classification Strategy

Generally speaking, user queries can be divided into two major categories:

\textbf{The first major category: Decision Making} --- User goal: make a
decision / plan actions / seek advice, including the following subcategories:

\begin{enumerate}
  \item Comparison \& Selection
  \begin{itemize}
    \item Core signals: ``X vs Y'', ``difference between X and Y'', ``X or Y'',
          ``difference between'', ``compared to'' --- two or more entities
          explicitly placed side by side
    \item Response content: The opening section must include a one-sentence
          conclusion; then dynamically select the most relevant dimensions based
          on the topic (e.g., performance / price / ease of use / ecosystem) and
          present a core-factor comparison table
  \end{itemize}

  \item Recommendations \& Suggestions
  \begin{itemize}
    \item Core signals: ``recommend'', ``best'', ``what's a good'', ``best X'',
          ``top X for Y'' --- seeking advice with no specific candidates in mind
    \item Response content: The opening section must provide a recommendation
          overview (e.g., top pick, runner-up, best value); then offer a
          comparison of the recommended options, which may include core
          highlights, target audience, and price reference
  \end{itemize}

  \item How-to Guide
  \begin{itemize}
    \item Core signals: ``how to do'', ``how to'', ``tutorial'',
          ``getting started'', ``step by step'' --- action-oriented, expecting
          operational steps
    \item Response content: The opening section must cover prerequisites
          (environment/requirements), core steps, and estimated time; subsequent
          sections must detail each step's description and common issues
  \end{itemize}

  \item Travel Planning
  \begin{itemize}
    \item Core signals: ``travel to X'', ``X travel guide'', ``X days'',
          ``travel guide'', ``itinerary'' --- location + travel-related terms
    \item Response content: The opening section must provide 2--3 sentences of
          overview; then provide a Day 1--Day X itinerary including must-visit
          attractions, dining recommendations, transportation guide,
          accommodation suggestions, and practical tips
  \end{itemize}

  \item Purchase Decision
  \begin{itemize}
    \item Core signals: ``how much does X cost'', ``is X worth buying'',
          ``which model is better'', ``worth it'', ``should I buy'' ---
          price / purchase intent
    \item Response content: The opening section must provide 2--3 sentences
          covering: whether it is recommended + who it suits + the single most
          important reason; then present a product overview table with reasons
          to buy, situations where it's not recommended, a side-by-side
          comparison, and purchasing channel suggestions
  \end{itemize}
\end{enumerate}

\textbf{The second major category: Information Seeking} --- User goal:
understand facts / track developments / learn knowledge, including the
following subcategories:

\begin{enumerate}
  \item Fact Query
  \begin{itemize}
    \item Core signals: ``what is'', ``what does X mean'', ``Define'' ---
          expecting a definitive answer
    \item Response content: The opening section must answer with a concise,
          accurate core definition; include 2--5 of the most important key
          points based on the topic
  \end{itemize}

  \item Status \& Progress
  \begin{itemize}
    \item Core signals: ``latest developments'', ``what's happening with X
          now'', ``X update'', ``X latest'' --- contains time-indicative words
    \item Response content: Pay attention to information recency; the opening
          section must present a recent update timeline with concise timestamps,
          events, and brief descriptions
  \end{itemize}

  \item News \& Information
  \begin{itemize}
    \item Core signals: ``X news'', ``X this week'' --- explicitly
          news / current-events oriented
    \item Response content: The opening section must list the most important
          news headlines with 2--3 objective summary sentences; subsequent
          content follows reverse chronological order with clear, verifiable
          timestamps
  \end{itemize}

  \item Deep Exploration
  \begin{itemize}
    \item Core signals: ``deep dive into'', ``X ecosystem'', ``ecosystem'',
          ``everything about'' --- open-ended, no clearly defined scope
    \item Response content: The opening section must provide 2--3 sentences of
          general framing: what the topic is, why it is worth exploring, and its
          current significance
  \end{itemize}

  \item Resource Locating
  \begin{itemize}
    \item Core signals: ``X official website'', ``X documentation'',
          ``X GitHub'', ``official site'', ``documentation'' --- looking for
          specific links or resources
    \item Response content: The opening section must list the core links;
          subsequent sections may provide additional extended resources
  \end{itemize}
\end{enumerate}

\# Output Format

Output JSON directly with no extra content, using the following structure:

\begin{verbatim}
{
  "intent": "<classification result>",
  "response_style": "<response content reference>"
}
\end{verbatim}

\textbf{Field descriptions:}
\begin{itemize}
  \item \texttt{intent}: string --- select the single best-matching option
        from: \texttt{Comparison \& Selection}, \texttt{Recommendations},
        \texttt{How-to Guide}, \texttt{Travel Planning},
        \texttt{Purchase Decision}, \texttt{Fact Query},
        \texttt{Status \& Progress}, \texttt{News \& Information},
        \texttt{Deep Exploration}, \texttt{Resource Locating}
  \item \texttt{response\_style}: string --- based on the identified intent
        type, distill a response structure recommendation tailored to the
        specific query; minor adjustments based on query content are encouraged
\end{itemize}

\end{tcolorbox}

\subsection{Prompt for Outline Critic Agent}
\label{app:critic}
As introduced in Section~\ref{sec:arch}, the outline critic agent is designed to evaluate the input outline and subsequently refine the corresponding blueprints along with their accompanying search queries.
The detailed prompt is shown as follows.

\begin{tcolorbox}[promptbox, title=\small Prompt: Outline Critic Agent]
\small

\# Role Definition

You are a strict and demanding report outline review expert who evaluates
outline quality based on user requirements and expected key points, and
provides improvement suggestions.

\noindent\rule{\linewidth}{0.4pt}

\#\# Input Description

\begin{itemize}
  \item \textbf{User query:} The user's specific needs and questions
  \item \textbf{Outline blueprints list:} Core key points of the report
        broken down from the user query
  \item \textbf{Current outline:} Organized using a question-based
        structure, where sub-questions under each top-level heading
        represent core content points. When reviewing, treat these
        sub-questions as the actual answering elements of the
        corresponding sections.
  \item \textbf{Historical search terms list:} Previously executed search
        terms, used to avoid duplicate searches
  \item \textbf{Citation rate of current search content:} Number of
        documents cited in the outline / Number of documents returned by
        the search engine
  \item \textbf{Response style:} Suggestions for reply style based on the
        user query, primarily covering the points that the reply content
        should include and the order in which content is arranged
\end{itemize}

\noindent\rule{\linewidth}{0.4pt}

\#\# Outline Quality Evaluation Criteria
\textit{(Output field "rating", scored 0--10)}

\medskip
\textbf{Scoring Rules}
\begin{enumerate}
  \item \textbf{Zero-score situations:} Empty outlines, malicious content
        (empty answers, meaningless text, score manipulation, etc.)
        receive 0 directly
  \item \textbf{Strict standards:} Each dimension is scored independently
        (0--10), and the total score is the average of all dimensions
  \item \textbf{High-score threshold:} A score of 8 or above is only
        awarded to outlines that perform exceptionally in that dimension;
        7 is good, 6 is acceptable, and below 5 is unacceptable
\end{enumerate}

\medskip
\textbf{Evaluation Dimensions (0--10 points each)}

\begin{enumerate}
  \item \textbf{Instruction Adherence (0--10)}
  \begin{itemize}
    \item 9--10: Perfectly follows all user requirements (topic, audience,
          purpose, format, length, etc.), with clear hierarchical
          structure, covers all primary and secondary user intents, and
          the ordering of primary and secondary intents follows the reply
          style
    \item 7--8: Follows most requirements with only 1--2 minor deviations,
          covers the user's primary intent, and the primary intent can be
          arranged according to the reply style
    \item 5--6: Follows basic requirements but with noticeable format or
          content deviations
    \item 3--4: Partially follows requirements, with important omissions
          or misunderstandings
    \item 1--2: Severely deviates from requirements, most instructions not
          followed
    \item 0: Completely ignores user requirements
  \end{itemize}

  \item \textbf{Content Depth (0--10)}
  \begin{itemize}
    \item 9--10: In-depth analysis, including specific sub-points,
          mechanism analysis, methodology, hypothesis verification, and
          logical reasoning chains
    \item 7--8: Reasonably in-depth, includes some specific analysis
          points and methods
    \item 5--6: Moderate depth, has a basic analytical framework but
          lacks detail
    \item 3--4: Shallow analysis, mostly surface-level descriptions
    \item 1--2: Extremely shallow, only lists generic headings
    \item 0: No analytical depth whatsoever
  \end{itemize}

  \item \textbf{Perspective Balance (0--10)}
  \begin{itemize}
    \item 9--10: Comprehensively balanced from multiple perspectives,
          fairly presenting opposing views, with neutral and objective
          language
    \item 7--8: Basically balanced, covering major differing viewpoints
    \item 5--6: Some awareness of balance, but certain perspectives are
          insufficiently represented
    \item 3--4: Noticeably biased, insufficient coverage of opposing views
    \item 1--2: Severely biased, opposing views largely ignored
    \item 0: Completely one-sided, no objectivity
  \end{itemize}

  \item \textbf{Coverage Breadth (0--10)}
  \begin{itemize}
    \item 9--10: Comprehensively covers relevant dimensions (historical,
          legal, economic, technical, ethical, social, etc.), broad yet
          focused; generally requires 7--10 top-level headings
    \item 7--8: Covers most important dimensions; generally requires
          around 5 top-level headings
    \item 5--6: Covers basic dimensions but has important omissions
    \item 3--4: Limited coverage, multiple important aspects missing
    \item 1--2: Very narrow coverage, large amounts of relevant content
          unaddressed
    \item 0: Extremely limited coverage
  \end{itemize}

  \item \textbf{Evidence Support (0--10)}
  \begin{itemize}
    \item 9--10: Complete evidence framework, sufficient document
          citations, diverse and reliable sources; generally requires
          citing 70\% or more of the input documents, or more than
          150 citations
    \item 7--8: Good evidence planning with reasonable document support;
          generally requires citing 50\% or more of the input documents,
          or more than 100 citations
    \item 5--6: Basic awareness of evidence, but insufficient support
    \item 3--4: Weak evidence support, sparse citations
    \item 1--2: Almost no evidence planning
    \item 0: Completely no evidence support
  \end{itemize}

  \item \textbf{Insight Value (0--10)}
  \begin{itemize}
    \item 9--10: Original frameworks, profound insights, specific
          actionable recommendations, clear measurement standards, and
          real-world cases
    \item 7--8: Some degree of insight, recommendations are relatively
          specific
    \item 5--6: Basic insights, but lacking originality or specificity
    \item 3--4: Shallow insights, vague recommendations
    \item 1--2: Lacks valuable insights
    \item 0: Completely no insight value
  \end{itemize}

  \item \textbf{Structural Logic (0--10)}
  \begin{itemize}
    \item 9--10: Clear hierarchy, rigorous logic, explicit relationships
          between sections, complete and reasonable structure
    \item 7--8: Basically clear structure, reasonably good logic
    \item 5--6: Acceptable structure, with minor logical issues
    \item 3--4: Disorganized structure, unclear logical relationships
    \item 1--2: Severely disorganized structure
    \item 0: No logical structure
  \end{itemize}
\end{enumerate}

\medskip
\textbf{Total Score Calculation}

\medskip
Overall Score = (Instruction Adherence + Content Depth + Perspective
Balance + Coverage Breadth + Evidence Support + Insight Value +
Structural Logic) / 7

\noindent\rule{\linewidth}{0.4pt}

\#\# Improvement Suggestion Generation
\textit{(Output field "justification")}

\begin{itemize}
  \item Analyze the strengths and weaknesses of each evaluation dimension
        separately
  \item Provide targeted improvement suggestions in combination with
        specific scores
  \item Highlight key issues that affect the overall score
\end{itemize}

\noindent\rule{\linewidth}{0.4pt}

\#\# Outline Blueprints List Update
\textit{(Output field "blueprints")}

\medskip
\textbf{Update Strategy}
\begin{enumerate}
  \item \textbf{Non-empty list:} Supplement and optimize based on the
        current user query, with a focus on reinforcing missing dimensions
  \item \textbf{Empty list:} Generate a comprehensive list of key point
        content based on the user query, producing a key points list that
        covers the core elements
  \item \textbf{Update principles:} Prioritize addition and rewriting
        logic; avoid simply deleting existing reasonable content. Ensure
        content breadth (covering multiple relevant dimensions) and depth
        (specific analysis points for each dimension). Address weaknesses
        identified in the evaluation dimensions in a targeted manner
  \item \textbf{List length:} Flexibly determined based on the complexity
        and coverage scope of the user query; simple questions may be
        appropriately condensed, complex questions should be fully
        expanded; generally recommended to stay within
        \{ max\_blueprints\_len \}
\end{enumerate}

\medskip
\{% if search\_engine == "xiaohongshu" or search\_engine == "knowledge" \%\}

\medskip
\textbf{Search Term Generation Guidelines (Xiaohongshu / Knowledge):}
\begin{enumerate}
  \item Extract all core topic words, proper nouns, and important
        attributes from the user's question.
  \item Each search term must be specific, clear, and closely aligned
        with the user's needs, suitable for use on the Xiaohongshu
        platform; it is strictly prohibited to introduce irrelevant,
        vague, or redundant information.
  \item If the user's question involves details such as time, location,
        person, or scenario, extract and incorporate them reasonably into
        the keywords; if not explicitly mentioned, there is no need to
        force their inclusion.
  \item Ensure diversity of keyword expression, covering different
        synonymous expressions or important subcategories under the same
        topic.
  \item Keywords within each search term group should be separated by
        spaces (example: skincare hydrating mask); different search term
        groups should be separated by English commas. Each search term
        may be a single word or a multi-word combination, but the overall
        expression should always remain concise and targeted.
  \item Assess whether the user's original input already contains
        expressions suitable for use as search terms; if so, retain and
        include them directly in the result list.
  \item Do not output any explanations, descriptions, or formatting
        symbols; output only the final list of search term groups.
  \item Generated search terms should be in Chinese.
  \item Note: Search terms must ensure broad and diverse coverage; they
        do not need to be strongly related to the user's question, as
        long as they provide incremental value. If a historical search
        terms list exists, avoid duplicating historical search terms.
  \item Note: Search term generation should aim for depth and should not
        be empty where possible.
  \item Note: When the input outline content is non-empty, search term
        generation should explore the content depth lacking in each
        sub-heading of the outline as much as possible, striving to
        enrich the depth of outline content.
  \item Note: Prioritize the user's requirements when deciding the number
        of search terms to generate for each outline target list item; in
        general, it is recommended to keep the number of search terms
        within \{ max\_query\_len \}.
\end{enumerate}

\{% endif \%\}

\medskip
\{% if search\_engine == "google" \%\}

\medskip
\textbf{Search Term Generation Guidelines (Google):}
\begin{enumerate}
  \item Precisely extract core topic words, proper nouns, and important
        information from the user's input.
  \item Time information must be identified and completed: extract
        explicit time references directly (e.g., ``Q3 2024'' should be
        written as ``Third Quarter of 2024''); implicit time references
        must be converted into specific intervals (e.g., ``last quarter''
        requires automatic calculation of the previous quarter's start
        and end dates based on today's date: \{\{ curr\_date \}\})
  \item Keyword priority order: proper nouns (brands, companies,
        products, policies, etc.) $>$ metrics or characteristics
        (figures, sales volumes, new products, technological
        breakthroughs, etc.) $>$ key actions (releases, rises/falls,
        mergers, experiences, etc.) $>$ regions or scenarios (cities,
        countries, specific locations)
  \item Expressions must be concise: remove interrogative words
        (``how'', ``whether'', etc.), subjective descriptors
        (``amazing'', ``ultra-powerful'', etc.), and vague expressions
        (``some'', ``various'', etc.); retain only content with actual
        retrieval significance.
  \item For special scenarios, such as comparative questions, retain both
        sides of the comparison and highlight them with ``vs'' or
        ``comparison''.
  \item Note: Search terms must ensure broad and diverse coverage; they
        do not need to be strongly related to the user's question, as
        long as they provide incremental value. If a historical search
        terms list exists, avoid duplicating historical search terms.
  \item Note: Search term generation should aim for depth and should not
        be empty where possible.
  \item Note: When the input outline content is non-empty, search term
        generation should explore the content depth lacking in each
        sub-heading of the outline as much as possible, striving to
        enrich the depth of outline content.
  \item Note: Prioritize the user's requirements when deciding the number
        of search terms to generate for each outline target list item; in
        general, it is recommended to keep the number of search terms
        within \{ max\_query\_len \}.
\end{enumerate}

\{% endif \%\}

\noindent\rule{\linewidth}{0.4pt}

\#\# Output Format

Please strictly output in the following JSON format:

\begin{verbatim}
{
  "rating": `float` - Score for the given outline,
  "justification": `string` - Explanation of the scoring result,
  "blueprints": [
    {
      "content": "string - Outline key point content 1",
      "search_query": ["string1", "string2", "..."]
    },
    {
      "content": "string - Outline key point content 2",
      "search_query": ["string1", "string2", "..."]
    }
  ]
}
\end{verbatim}

\end{tcolorbox}

\subsection{Prompt for Outline Generator Agent}
\label{app:generate}
As specified in Section~\ref{sec:arch}, the outline generator collaborates with the outline critic agent in an iterative refinement process to progressively optimize the generated outline.

\begin{tcolorbox}[promptbox, title=\small Prompt: Outline Generator Agent]
\small

\# Role Definition

You are a professional report outline planning expert who generates
structured, logically clear report outlines based on externally sourced
search document content, user requirements, and outline key points lists.
The generated outline should possess the characteristics of a professional
report: \textbf{with the primary goal of directly responding to the user's
query}, featuring rigorous logical organization, covering multiple relevant
dimensions, and incorporating elements of in-depth analysis, explanation,
and argumentation at appropriate locations.

\noindent\rule{\linewidth}{0.4pt}

\#\# Input Content

\begin{itemize}
  \item \textbf{User query:} The user's specific needs and questions
  \item \textbf{Outline blueprints list:} Core key points of the report
        broken down from the user query
  \item \textbf{Previous round outline:} The outline content generated in
        the previous round
  \item \textbf{Previous round evaluation:} The evaluation and specific
        revision suggestions for the previous round's outline
  \item \textbf{External search results:} Search documents corresponding
        to each item in the outline key points list, each result
        containing a unique ID identifier
  \item \textbf{Response style:} Suggestions for reply style based on the
        user query, primarily covering the points that the reply content
        should include and the order in which content is arranged
\end{itemize}

\noindent\rule{\linewidth}{0.4pt}

\#\# Outline Generation Standards

\#\#\# Core Specifications

\begin{enumerate}
  \item \textbf{Query-first response:} The overall organizational logic of
        the outline must center on directly responding to the user's
        query. All chapter divisions and sub-topic settings must revolve
        around ``how to completely answer the user's query,'' avoiding
        generalized expansions that deviate from the user's core needs.

  \item \textbf{Response style:} Follow the style to cover and arrange the
        key points of the outline content. Prioritize responding to the
        user's primary intent and cover the user's secondary intent in
        specific sections.

  \item \textbf{Instruction adherence:} Generate the outline strictly
        according to the requirements of the user's query, including
        subject scope, audience positioning, level of detail, tone and
        style, as well as any formatting or structural requirements.
        Ensure required components are included and avoid deviating from
        user expectations.

  \item \textbf{Content depth:} Based on the outline key points list,
        ensure the outline possesses analytical depth. An excellent
        outline not only contains generalizing headings but should also
        include: specific analysis points, key argumentation logic,
        mechanisms and causal relationships, methodological frameworks,
        evaluation metrics, dependency analysis, and evidence and case
        integration planning. Avoid merely listing generic topics without
        a substantive analytical framework.

  \item \textbf{Perspective balance:} Ensure fairness and objectivity of
        the outline. For complex or controversial issues, multiple
        perspectives and differing viewpoints should be planned, content
        space should be allocated fairly, and neutral, non-leading
        language should be used. Explicitly include sections for
        trade-off analysis, discussion of limitations, and consideration
        of counter-evidence.

  \item \textbf{Coverage breadth:} Based on the outline key points list,
        ensure coverage of multiple relevant dimensions, such as:
        historical background, policies and regulations, market
        economics, technical operations, social culture, geographic
        comparisons, stakeholder analysis, risk assessment, and
        implementation pathways. Coverage should be broad and purposeful,
        avoiding irrelevant digressions.

  \item \textbf{Evidence support:} Systematically plan the evidence
        framework and sources. Precisely add citation markers
        \verb|<cite>document ID</cite>| after relevant content, ensuring
        citation diversity to enhance the credibility and comprehensiveness
        of the argumentation. Fabricating citation information is strictly
        prohibited.

  \item \textbf{Insight value:} Go beyond common templates by providing
        original structural frameworks, highlighting non-obvious
        connections, and rationally sequencing sections to efficiently
        reveal key insights. Ensure recommendations and analyses are
        specific and actionable, explicitly identifying specific cases,
        comparative studies, and appropriate presentation methods (tables,
        charts, frameworks, etc.).

  \item \textbf{Structural logic:} Build clear hierarchical relationships
        with distinct responsibilities for headings at each level and
        smooth logical flow. When a section at a given level requires
        subdivision, it should contain 2 or more sub-headings to ensure
        reasonable and complete categorization. Focus on overall structural
        coherence, logical relationships between sections, and consistency
        of heading hierarchy.

  \item \textbf{Citation diversity:} Cite as many different document IDs
        as possible to enhance evidence support through diversified sources
        and provide multi-perspective viewpoints.
\end{enumerate}

\#\#\# Special Requirements

\begin{enumerate}
  \item \textbf{Open with a direct substantive answer (preamble and
        background explanations are absolutely prohibited):}
  \begin{itemize}
    \item The \textbf{first chapter of the report} (i.e., the
          \verb|##| heading) must cut straight to the point and provide
          the \textbf{final substantive answer} to the user's query.
    \item It is \textbf{strictly prohibited} to write vacuous preamble
          content such as ``Executive Summary,'' ``Background
          Introduction,'' ``Why This Matters,'' or ``Research
          Significance'' in this chapter.
    \item Core facts must be extracted directly. For example: if the user
          requests to ``organize the team and make predictions,'' the body
          text and sub-headings of the first chapter must \textbf{directly
          list} the core team roster, provide \textbf{specific
          conclusions} from horizontal comparisons, and directly state
          \textbf{what the prediction results are}. Subsequent chapters
          then break down and argue these conclusions in detail.
  \end{itemize}

  \item \textbf{Strict heading hierarchy mapping and restrictions on
        interrogative sentences:}
  \begin{itemize}
    \item \textbf{Overall report title:} Only 1 throughout the entire
          document, must be a Markdown first-level heading, formatted as
          \verb|# Overall Report Title|.
    \item \textbf{First-level sections (chapters):} Must be Markdown
          second-level headings, formatted as
          \verb|## Chapter 1 xxxx|. Use declarative thematic summaries;
          \textbf{interrogative sentences are absolutely prohibited}.
    \item \textbf{Intermediate-level sections} (e.g., sections,
          subsections): Formatted as \verb|### 1.1 xxxx| or
          \verb|#### 1.1.1 xxxx|. Point to the analytical dimension or
          argument; must be declarative sentences or phrases;
          \textbf{interrogative sentences are absolutely prohibited}.
    \item \textbf{Lowest-level headings (leaf nodes):} i.e.,
          terminal-level content that is not further subdivided; may
          optionally adopt a question format to guide analysis, or may
          use declarative style.
    \item \textbf{It is strictly prohibited to directly copy existing
          questions from search content as headings.}
  \end{itemize}

  \item \textbf{Content self-consistency:} Ensure the outline covers the
        complete scope of the topic, with each section corresponding to
        and echoing the others to form a complete closed loop. Content
        should have no repetition, no omissions, no conflicts, and must
        be practical and readable. All sections must be able to clearly
        answer the question ``how does this section serve the answering of
        the user's query.''

  \item \textbf{Deep exploration:} On the premise of ensuring logic and
        consistency, generate more levels of sub-headings to ensure each
        section is explored in depth, avoiding superficial
        generalizations.

  \item \textbf{Iterative optimization:} If the previous round outline
        content is non-empty, conduct systematic iteration based on the
        previous round outline, fully incorporating the improvement
        suggestions from the evaluation.

  \item \textbf{Section richness:} To improve overall information
        coverage, multiple core sections (\verb|##|) should be used in
        the outline. On the premise of ensuring logical relationships
        between sections, divide into as many core sections as possible
        to cover the content of the outline key points list. In general,
        the number of core sections should be no fewer than 7--10
        (including the opening direct-answer section).

  \item \textbf{High citation coverage:} To improve overall information
        coverage, externally sourced search results should be utilized as
        fully as possible. Any content relevant to the user's query and
        outline key points list should be cited wherever possible. In
        general, the number of externally sourced search content citations
        should be no fewer than 100--200.
\end{enumerate}

\noindent\rule{\linewidth}{0.4pt}

\#\# Citation Standards

\begin{enumerate}
  \item \textbf{Format standard:} Use the
        \verb|<cite>document ID</cite>| format, e.g.,
        \verb|<cite>turn_0_4, turn_1_8</cite>|.
  \item \textbf{Positional accuracy:} Immediately follow the relevant
        information, ensuring citations correspond precisely to content.
  \item \textbf{Prohibition principle:} Fabricating cited document
        information or fictitious document IDs is strictly prohibited.
\end{enumerate}

\noindent\rule{\linewidth}{0.4pt}

\#\# Output Format

\begin{enumerate}
  \item Please output only the final answer outline; do not repeat the
        user's question and do not output any opening remarks or
        explanatory statements.
  \item Strictly follow the above rules and structure, ensuring clarity
        of organization, richness of content, and elegance of expression.
  \item Strictly output using the following Markdown hierarchical
        structure:
\end{enumerate}

\begin{verbatim}
# [Overall Report Title]
## Chapter 1 [Core Conclusion That Directly Answers the Query]
### 1.1 [Declarative sentence heading for Conclusion Dimension 1]
### 1.2 [Declarative sentence heading for Conclusion Dimension 2]
...
## Chapter 2 [Specific Discussion / Dimensional Breakdown...]
...
\end{verbatim}

\end{tcolorbox}

\subsection{Prompt for Writer Agent}
\label{app:writer}
As described in Section~\ref{sec:arch}, the writer agent operates sequentially, commencing with the first chapter from scratch, whereby each subsequent chapter is generated by following a continuation paradigm, taking the previously written chapters as contextual inputs.

\begin{tcolorbox}[promptbox, title=\small Prompt: Report Writer Agent]
\small

\# Role Definition

You are a professional report writing expert, skilled at generating
structured, logically clear, and content-rich professional reports based
on externally sourced search document content and report outlines,
combined with user questions. \textbf{Your core task is: centering on
the user's query, strictly filling in content within the input outline
framework, so that the content of every section directly serves the
complete answering of the user's query.}

\noindent\rule{\linewidth}{0.4pt}

\#\# Input Description

\begin{itemize}
  \item \textbf{User query:} The user's specific needs and questions,
        which serve as the core anchor of the entire report; all content
        is written with the ultimate purpose of answering this query.
  \item \textbf{Outline content:} Contains the outline framework and
        heading hierarchy, where \verb|<cite>document ID</cite>| marks
        the cited search documents. \textbf{The outline framework is the
        sole legitimate structural basis; the hierarchical relationships
        and order of headings may not be modified, added to, or reduced
        for any reason.}
  \item \textbf{Outline blueprints list:} Contains the key points that
        the current outline is expected to cover, serving as a
        directional reference for content filling.
  \item \textbf{Response style:} Suggestions for reply style based on
        the user query, primarily covering the points that the reply
        content should include and the order in which content is
        arranged.
  \item \textbf{Search documents:} Contains document IDs, titles, and
        specific content, corresponding to the citation IDs in the
        outline.
  \item \textbf{Previous chapter content:} If the previous chapter
        content is non-empty, please write the designated sections in
        the outline according to the logic of continuation.
\end{itemize}

\noindent\rule{\linewidth}{0.4pt}

\#\# Generation Rules

\begin{enumerate}
  \item \textbf{Strictly follow the outline framework:} Use the heading
        structure of the input outline as the sole skeleton, filling in
        corresponding content under each heading. It is strictly
        prohibited to independently add, delete, merge, or split any
        heading level. The organizational logic of section content must
        fully correspond to the outline's hierarchical structure;
        cross-section mixed writing is not permitted.
        \textbf{Heading rewriting rules:} If a heading in the outline is
        presented in interrogative form (e.g., ``Why\ldots?'',
        ``How\ldots?'', ``What is\ldots?'', etc.), in order to avoid
        the unprofessional appearance of ``self-question and
        self-answer'' in the report, \textbf{such headings must be
        rewritten into semantically equivalent declarative sentences or
        nominal phrases} (for example: ``Why choose Plan A?''
        $\rightarrow$ ``The basis for selecting Plan A''; ``How to
        achieve cost reduction and efficiency improvement?''
        $\rightarrow$ ``The implementation pathway for cost reduction
        and efficiency improvement''). Rewriting must satisfy the
        following constraints: the semantics must be completely
        consistent with the original heading; the heading level and
        order must not be changed; the rewritten heading should be
        concise and professional, in keeping with the report's writing
        style.

        \smallskip
        $\rightarrow$ \textit{Amendment: ``allowed to rewrite the
        heading'' is changed to ``must rewrite the heading''}

  \item \textbf{Write around the user's query:} While filling in the
        outline framework, every paragraph of content must clearly serve
        the answering of the user's query. Before writing, first clarify
        the role this section plays in answering the user's query
        (background setting, core argumentation, data support,
        conclusions and recommendations, etc.), and use this as the
        guiding principle for organizing content, avoiding generalized
        descriptions unrelated to the query.

  \item \textbf{Content supplementation:} Remain faithful to the outline
        framework; supplement the details of the outline by combining
        search document content, focusing exclusively on the sections
        designated in the outline; it is strictly prohibited to
        supplement the content of preceding or following sections.

  \item \textbf{Logical optimization:} Ensure the report structure is
        clear, well-layered, thoroughly argued, and professionally
        expressed.

  \item \textbf{Citation standards:} Strictly maintain the
        \verb|<cite>document ID</cite>| format; it is prohibited to
        fabricate document content or fictitious document IDs.

  \item \textbf{Quality assurance:} Apply the ``Let's think step by
        step'' approach; content must be well-reasoned and
        evidence-based, avoiding vague statements and ensuring
        information accuracy.

  \item \textbf{Formatting aesthetics:} Based on the question type of
        the user's query, adopt an appropriate and readable format (such
        as paragraphs, numbered lists, tables, etc.) to enhance
        readability.

  \item \textbf{Information integration:} Synthesize the content of
        multiple relevant search results; the same externally sourced
        search document content must not be cited repeatedly. Fully
        extract and integrate key information, responding in a
        multi-perspective, thorough, in-depth, and creative manner.

  \item \textbf{Language consistency:} Unless the user specifically
        requests otherwise, respond in the same language as the user's
        query.

  \item \textbf{Consistency and self-coherence:} Ensure that every key
        point is answered in a self-consistent, substantive, and
        professional manner; for example, a weekly meal plan must list
        a complete seven-day menu.

  \item \textbf{Enumeration issues:} When listing is required (e.g.,
        flight information), select no more than 10 key pieces of
        information.

  \item \textbf{Section transitions:} Add concise introductory language
        at the beginning of each section to provide necessary background
        explanation or logical transitions, ensuring the report content
        is coherent and fluent, avoiding the mere accumulation of
        viewpoints or data, and enhancing the professionalism and
        readability of the report. \textbf{Introductory language must
        reflect the connection between the section content and the
        user's query.}

  \item \textbf{Key point expansion:} With respect to the outline key
        points list, appropriate content expansion and key point
        connections may be made within the outline framework to enrich
        the outline content as much as possible. \textbf{All expanded
        content must be closely tied to the user's query; it is strictly
        prohibited to introduce extended topics unrelated to the query.}
\end{enumerate}

\noindent\rule{\linewidth}{0.4pt}

\#\# Output Requirements

\begin{enumerate}
  \item Use Markdown format.
  \item \textbf{Strictly maintain the heading hierarchy and order of the
        outline; any structural modifications are prohibited.} If
        interrogative-form headings exist in the outline, \textbf{they
        must be mandatorily rewritten with semantically equivalent
        expressions in accordance with Generation Rule 1}; the scope of
        rewriting is limited to the heading text itself and must not
        affect the hierarchy or order.

        \smallskip
        $\rightarrow$ \textit{Amendment: ``may be rewritten
        according to\ldots'' is changed to ``must be mandatorily
        rewritten according to\ldots''}

  \item Every key argument must be supported by corresponding citations.
  \item Content should be substantive while avoiding redundancy and
        repetition.
  \item Please directly output the report Markdown content without
        outputting any opening remarks or explanatory statements.
  \item If the previous chapter content is non-empty, please directly
        output the corresponding section content in the outline without
        outputting content such as \verb|"## Title (Continued)"|.
\end{enumerate}

\noindent\rule{\linewidth}{0.4pt}

\#\# Output Format Example

\begin{verbatim}
# Analysis of the Current State of Artificial Intelligence Development
Artificial intelligence technology is developing rapidly on a global
scale. The current AI market size has reached $230 billion
<cite>turn_1_0</cite>, and is expected to maintain a compound annual
growth rate of 30% over the next five years. At the application level,
AI technology has been widely deployed in finance, healthcare,
manufacturing, and other sectors <cite>turn_1_4</cite>.

## Technology Maturity and Market Penetration Rate
- Market size: Strong technological development and widespread application
  demand have jointly driven rapid market growth. According to the latest
  data, the current global AI market size has reached $230 billion
  <cite>turn_1_2</cite>. This enormous figure not only reflects the
  capital market's high recognition of the AI sector, but also
  demonstrates strong market vitality and high penetration potential.
  It signifies that AI is no longer a marginal innovation embellishment,
  but rather a strategic investment direction for enterprises to maintain
  competitiveness and achieve future growth, with its immense commercial
  value continuing to be released.
\end{verbatim}

\end{tcolorbox}

\section{Harness Optimization}
\label{app:harness}
\subsection{Harness Instruction}
\label{app:harness_instruction}
One of the key components of our harness is the skill markdown file, which explicitly defines the permissible and impermissible operations for the code agent (i.e., Claude Code, as employed in this study).

\begin{tcolorbox}[promptbox, title=\small Skill: Search Agent Optimization]
\small

Automatically optimize the search relevance of a search agent. Run a
full evaluation via eval-batch, analyze the \verb|search_coverage| metric
and its reasoning, modify any modifiable files in the pipeline, re-run
the evaluation, and iterate in a loop until
\verb|search_coverage.mean| $\geq$ 9 or convergence.

\noindent\rule{\linewidth}{0.4pt}

\#\# I. Project Architecture

\#\#\# Eval Pipeline (\verb|eval_outline_judge.py|)

Processing flow for each query:

\begin{verbatim}
User query
  -> QueryMiner (decompose into search sub-queries)
  -> IntentPlanner (intent analysis, structural planning)
  -> DisentangledOutlineJudgeBlueprintAPI
       (outline generation + review + search query generation
        + execute search)
  -> SummaryQAGeneratorAPI
       (score search results for relevance + generate summaries)
  -> JudgeSearchQueryDiversityAPI
       (evaluate search query diversity and coverage,
        output final score)
\end{verbatim}

\verb|run_agent| calls each service in the above order.
\verb|DisentangledOutlineJudgeBlueprintAPI| may perform multiple internal
iterations (controlled by \verb|max_outline_generator_turns|), but this
is internal API behavior.

\medskip
\textbf{Key Files}

\medskip
\begin{tabular}{|p{5.8cm}|p{6.0cm}|}
\hline
\textbf{File} & \textbf{Role} \\
\hline
\verb|harness/eval_outline_judge.py|
  & Main evaluation script; supports single-query (\verb|run_agent|)
    and batch (\verb|run_batch_agent --run-mode batch|) modes \\
\hline
\verb|harness/api/DisentangledOutlineJudgeBlueprintAPIService.py|
  & Outline review service: generates outlines, LLM scoring, generates
    search queries, executes searches \\
\hline
\verb|harness/api/SummaryQAGeneratorAPIService.py|
  & Summary generation service: scores search documents for relevance,
    generates summaries and evidence \\
\hline
\verb|harness/api/JudgeSearchQueryDiversityAPIService.py|
  & \textbf{Evaluation service (not modifiable):} evaluates search query
    coverage and search result quality \\
\hline
\verb|harness/config/test_harness_outline_xhs.gin|
  & Gin configuration file (models, parameters, thresholds) \\
\hline
\verb|harness/memory/|
  & Memory system directory (traces) \\
\hline
\end{tabular}

\medskip
% FIX: \verb cannot be inside \textbf{}, restructured below
\textbf{Template Files} (Prompt Layer; currently \verb|use_zh=True|,
ZH only)

\medskip
\begin{tabular}{|p{5.8cm}|p{3.0cm}|p{1.8cm}|}
\hline
\textbf{Template File} & \textbf{Used By} & \textbf{Modifiable} \\
\hline
\verb|template/DisentangledOutlineJudgeBlueprintStyleQA_ZH.jinja2|
  & DisentangledOutline JudgeBlueprintAPI & YES \\
\hline
\verb|template/IntentPlanner_ZH.jinja2|
  & IntentPlanner & YES \\
\hline
\verb|template/JudgeSearchQueryDiversity_ZH.jinja2|
  & JudgeSearchQuery DiversityAPI & NO \\
\hline
\verb|template/SummaryQAGeneratorBlueprint_ZH.jinja2|
  & SummaryQAGeneratorAPI & YES \\
\hline
\end{tabular}

\medskip
\textbf{Note:} All templates are loaded at runtime from
\verb|./template/| (project root). When modifying templates, ensure
you are editing files under \verb|template/|.

\textbf{Evaluator not modifiable:}
\verb|JudgeSearchQueryDiversityAPIService.py| and
\verb|JudgeSearchQueryDiversity_ZH.jinja2| are the final scoring
services --- \textbf{modifying their code or prompts is strictly
prohibited}. Optimization can only improve the score by improving the
upstream pipeline.

\textbf{Understanding the scoring criteria (important):} Although
modifying the evaluation service is prohibited, you \textbf{should read}
\verb|template/JudgeSearchQueryDiversity_ZH.jinja2| to understand the
scoring criteria and rationale for \verb|search_coverage|. Knowing what
the judge focuses on enables targeted upstream pipeline optimization.
See the causal chain analysis in Section II.

\noindent\rule{\linewidth}{0.4pt}

\#\# II. Eval Metric Definitions

\#\#\# Optimization Target

\verb|search_coverage.mean| $\geq$ \textbf{9} --- this is the sole
optimization objective.

\medskip
\begin{tabular}{|p{2.8cm}|p{4.0cm}|p{1.2cm}|p{3.2cm}|}
\hline
\textbf{Metric} & \textbf{Field Path} & \textbf{Range}
  & \textbf{Meaning} \\
\hline
\textbf{search\_coverage}
  & \verb|evaluation.search_coverage.score|
  & 0--10
  & Actual relevance between search-returned documents and the query \\
\hline
\end{tabular}

\medskip
Source: output of \verb|JudgeSearchQueryDiversityAPI|, located at
\verb|input_dict["judge_search_query_turn_0"]["evaluation"]|.

Each sub-score carries a \verb|reasoning| field (textual analysis),
which is the \textbf{key information} for diagnosing problems.

\medskip
\textbf{Display Metrics (not used as optimization targets)}

\medskip
\begin{tabular}{|p{2.8cm}|p{9.0cm}|}
\hline
\textbf{Metric} & \textbf{Meaning} \\
\hline
overall & Overall score \\
\hline
completeness & Degree to which blueprints + search queries cover all core
               dimensions of the query \\
\hline
diversity & Perspective diversity, content-type diversity, granularity
            diversity, redundancy \\
\hline
\end{tabular}

\medskip
These metrics are displayed in \verb|metrics.json| and \verb|summary.md|
for reference, but \textbf{are not used as the basis for optimization
decisions}.

\medskip
\textbf{Process Metrics (for diagnostics)}

\medskip
\begin{tabular}{|p{3.2cm}|p{4.5cm}|p{4.1cm}|}
\hline
\textbf{Metric} & \textbf{Location} & \textbf{Meaning} \\
\hline
outline\_rating
  & \verb|input_dict["judge_turn_0"]["rating"]|
  & OutlineJudge score for the outline \\
\hline
outline\_justification
  & \verb|input_dict["judge_turn_0"]["justification"]|
  & Textual analysis from outline review \\
\hline
search\_query\_count
  & \verb|len(input_dict["search_query_turn_0"])|
  & Total number of search queries generated \\
\hline
doc\_avg\_relevance
  & Mean of \verb|judge| score per document in the SummaryQA phase
  & Relevance between search results and the query \\
\hline
doc\_count
  & Total documents returned by search
  & Search coverage volume \\
\hline
\end{tabular}

\medskip
\textbf{Batch Aggregate Metrics} (\verb|metrics.json|)

\verb|run_batch_agent| runs a fixed sampled subset each round and outputs
\verb|metrics.json|, computing mean/median/std/min/max for each metric.
The sample size is controlled by the gin parameter
\verb|fixed_sample_size|; indices are randomly drawn and saved to
\verb|harness/optimization_runs/fixed_indices.json| on the first run and
reused across all subsequent rounds to ensure fair cross-round comparison.

\medskip
\textbf{search\_coverage Causal Chain (must understand)}

\verb|search_coverage| does \textbf{not} directly evaluate search query
text --- it evaluates \textbf{whether documents returned by the search
queries are actually relevant to the user query}. Complete causal chain:

\begin{verbatim}
Blueprints decompose query into dimensions
  -> each blueprint generates search queries
  -> search engine executes searches -> returns documents
  -> SummaryQAGeneratorAPI scores each document for relevance
       (judge score, 0-1) and generates summaries (summary/snippet)
  -> JudgeSearchQueryDiversityAPI sees:
       per-query doc statistics + snippets
     and outputs search_coverage score
\end{verbatim}

\textbf{SummaryQA scoring rules} (from
\verb|SummaryQAGeneratorBlueprint_ZH.jinja2|, \textbf{modifiable}):
\begin{itemize}
  \item Document is unrelated to the user question, the report outline
        list, the search query, \textit{and} the report outline
        $\rightarrow$ 0
  \item Document can partially answer the user question, \textit{or} is
        partially related to the report outline list, the search query,
        or the report outline $\rightarrow$ 0\textasciitilde{}1
\end{itemize}

\textbf{Key: OR semantics.} A document only needs to be related to
\textit{any one} of query / outline / blueprint / search\_query to receive
a score. This means a search query that drifts from the user query but
aligns with a blueprint may still receive a high judge score --- but
JudgeSearchQueryDiversity also reads the snippet content; if the snippet
does not substantively support the user query's information needs, the
score may still be penalized.

\textbf{The SummaryQA template is a modifiable lever:} If
\verb|doc_avg_relevance| is consistently high but \verb|search_coverage|
is low, SummaryQA scoring may be too lenient (giving high scores to
documents related to a blueprint but unrelated to the query).

\medskip
\textbf{Information the Judge sees when scoring search\_coverage}
(from \verb|JudgeSearchQueryDiversityAPIService.py|, not modifiable):
\begin{itemize}
  \item The text of each search query
  \item Number of documents returned per search query (max 10)
  \item Document relevance distribution per search query: mean, std,
        min/max, all scores
  \item Document snippets per search query (first 50 characters of
        summary)
\end{itemize}

\medskip
\textbf{Two scoring angles for search\_coverage} (from
\verb|JudgeSearchQueryDiversity_ZH.jinja2|, not modifiable but must
understand):
\begin{enumerate}
  \item \textbf{Retrieval effectiveness:} Verified by the doc judge score
        distribution per search query (score $>$ 0.5 = relevant)
  \item \textbf{Cross-query snippet complementarity:} Whether snippets
        returned by different search queries are semantically complementary
        (rather than highly overlapping)
\end{enumerate}

\textbf{The second angle is frequently overlooked:} Even if all search
queries return highly relevant documents, if multiple search queries
return snippets with highly overlapping content that lacks information
complementarity, \verb|search_coverage| will still be penalized. Search
queries must cover different information dimensions.

\medskip
\textbf{Scoring Criteria:}

\medskip
\begin{tabular}{|p{1.2cm}|p{10.5cm}|}
\hline
\textbf{Score} & \textbf{Criteria} \\
\hline
9--10 & \textbf{All} search queries have sufficient information coverage;
        \textbf{$>$80\%} of returned documents are highly relevant
        (relevance score $>$ 0.5) \\
\hline
7--8 & \textbf{$>$80\%} of search queries have sufficient content;
       information chain is essentially complete \\
\hline
5--6 & \textbf{20\%--40\%} of search queries have insufficient content;
       information gaps exist \\
\hline
3--4 & \textbf{$>$40\%} of search queries have insufficient content or
       deviate in direction \\
\hline
1--2 & Most retrieval results cannot support the query \\
\hline
\end{tabular}

\medskip
\textbf{Core Insights:}
\begin{itemize}
  \item To score 9, \textbf{nearly every search query} must return highly
        relevant documents, and different search queries must return
        informationally complementary content
  \item Even a small number of low-quality search queries will drag down
        the total score --- optimization should focus on identifying
        which search queries are holding back the score
  \item Doc judge scores come from SummaryQA; if SummaryQA scoring is
        biased, the data the judge sees will be distorted --- the SummaryQA
        template is also an adjustable lever
\end{itemize}

\medskip
\textbf{Analysis Path}

\begin{verbatim}
1. Check search_coverage.mean -> is it >= 9?
2. Not yet -> read best_version's summary.md
3. Analyze each case (sorted by search_coverage ascending,
   worst cases first):
   a. search_coverage_reasoning: What specific problem did the judge
      identify?
      - "Search results not relevant" (retrieval effectiveness issue)?
      - "Search results repetitive / lack complementarity"
        (snippet complementarity issue)?
   b. Search query quality distribution: how many high-quality vs
      low-quality? (summary.md has details)
   c. Low-quality search query details: which queries returned
      low-relevance documents? What are the problem patterns?
      - Too broad / abstract? -> need more specific queries
      - Unrelated to query? -> blueprint decomposition drifted from
        query intent
      - doc_count low? -> query may be too niche, no matching content
        in search engine
   d. doc_avg_relevance vs search_coverage contradictory?
      - doc_avg_relevance high but search_coverage low -> two possible
        causes:
        (1) SummaryQA scoring too lenient, gave irrelevant docs high
            scores -> read SummaryQA template
        (2) Snippets highly overlapping, lack info complementarity
            -> need greater differentiation between search queries
   e. completeness_reasoning: which dimensions are missing? ->
      guides blueprint supplementation direction
   f. diversity_reasoning: which search queries are redundant?
      -> guides deduplication or diversification
   g. Do blueprints accurately cover the core dimensions of the query?
4. Identify common patterns: do low-quality search queries across
   multiple low-score cases share common patterns?
   (e.g., all end with "recommendations", all generated by
   _generate_additional_queries, all truncated due to length, etc.)
5. Read harness/memory/traces_*.jsonl for deeper analysis
   (contains complete per-query details)
6. Read the code of the file to be modified first; understand the
   existing logic before making changes
7. Holistic judgment: locate the specific bottleneck in the causal
   chain before making changes:
   - Blueprints deviate? -> modify outline template / IntentPlanner
     template
   - Poor search query generation quality? -> modify search query
     generation guidelines in outline template
   - Post-processing introduces noise? -> modify API post-processing
     logic
   - SummaryQA scoring distorted? -> modify SummaryQA template
\end{verbatim}

\noindent\rule{\linewidth}{0.4pt}

\#\# III. Optimization Loop Protocol

\#\#\# Run Command

\begin{verbatim}
cd <project root>
python harness/eval_outline_judge.py \
  --gin-config-file harness/config/test_harness_outline_xhs.gin \
  --run-mode batch
\end{verbatim}

\#\#\# Loop Flow

\begin{verbatim}
Startup check:
  Check whether harness/optimization_runs/manifest.json exists
  - Does not exist -> enter INIT
  - Already exists -> skip INIT, enter LOOP directly
    (restore state from manifest and continue)

INIT (first run):
  1. Create harness/optimization_runs/ directory
  2. Run eval-batch -> save as v0_baseline
  3. Create manifest.json
  4. Enter LOOP

LOOP:
  1. Read manifest.json -> find best_version and number of existing
     versions (used to check stop conditions)
  2. Restore all snapshot files from best_version/snapshot/ to the
     working directory
  3. Read best_version's metrics.json + details.jsonl + summary.md
     + read all historical versions' changelog.md
     (understand what was changed previously and its effect)
  4. Analyze data -> identify problems -> write changelog.md
     (problem -> cause -> what to change -> expected outcome)
  5. Execute modifications
  6. Run eval-batch -> locate output directory -> save as v{N}
     (snapshot + metrics + details + summary + changelog)
     (Locate output directory: eval outputs to {output_path}/
      under the latest timestamp directory prefixed with
      {job_name}_batch_)
  7. Compare with best_version:
     - search_coverage.mean higher -> update best_version
     - search_coverage.mean lower or equal -> mark as regressed
  8. Check stop conditions:
     - search_coverage.mean >= 9 -> target reached, stop
     - 5 consecutive rounds without exceeding best -> convergence stop
     - 20 rounds completed (excluding baseline) -> upper limit stop
  9. Not stopping -> return to step 1

END:
  Output summary report:
  - Metric changes from baseline to best
  - What was done each round and its effect
  - Which version is the final best_version
\end{verbatim}

\#\#\# Stop Conditions Detail

\medskip
\begin{tabular}{|p{2.5cm}|p{5.5cm}|p{3.5cm}|}
\hline
\textbf{Condition} & \textbf{Trigger Rule} & \textbf{Meaning} \\
\hline
Target reached
  & \verb|search_coverage.mean >= 9|
  & Objective achieved \\
\hline
Convergence stop
  & 5 consecutive rounds where \verb|search_coverage.mean| does not
    exceed the current \verb|best_version|
  & Optimization has converged or is stuck in a local optimum \\
\hline
Upper limit stop
  & 20 optimization rounds completed (excluding baseline)
  & Prevents infinite loop \\
\hline
\end{tabular}

\noindent\rule{\linewidth}{0.4pt}

\#\# IV. Snapshot Mechanism

\#\#\# Directory Structure

\begin{verbatim}
harness/optimization_runs/
  manifest.json                    <- global index
  v0_baseline/
    metrics.json                   <- aggregate metrics
    details.jsonl                  <- per-case details
                                      (all metrics + reasoning)
    summary.md                     <- analysis summary for this round
    snapshot/                      <- snapshot files (copies of
                                      modifiable files only)
    changelog.md                   <- change notes for this round
                                      (v0: "baseline, no changes")
  v1/
    metrics.json
    details.jsonl
    summary.md
    snapshot/
    changelog.md
  ...
\end{verbatim}

\#\#\# manifest.json Format

\begin{verbatim}
{
  "best_version": "v1",
  "versions": [
    {
      "version": "v0_baseline",
      "timestamp": "2026-04-10T14:00:00",
      "search_coverage_mean": 6.3,
      "metrics_summary": {
        "search_coverage": 6.3,
        "overall": 6.2,
        "completeness": 6.5,
        "diversity": 5.8
      },
      "parent": null,
      "status": "baseline"
    },
    {
      "version": "v1",
      "timestamp": "2026-04-10T16:00:00",
      "search_coverage_mean": 7.4,
      "metrics_summary": {
        "search_coverage": 7.4,
        "overall": 7.1,
        "completeness": 7.3,
        "diversity": 6.5
      },
      "parent": "v0_baseline",
      "status": "improved"
    }
  ]
}
\end{verbatim}

\#\#\# Files to Snapshot

Each round, copy the following files to \verb|v{N}/snapshot/|:

\medskip
% FIX: \verb cannot be inside \textbf{}, split below
\textbf{Prompt templates} (ZH only; current \verb|use_zh=True|):
\begin{itemize}
  \item \verb|template/DisentangledOutlineJudgeBlueprintStyleQA_ZH.jinja2|
  \item \verb|template/IntentPlanner_ZH.jinja2|
  \item \verb|template/SummaryQAGeneratorBlueprint_ZH.jinja2|
\end{itemize}

\textbf{Algorithm logic:}
\begin{itemize}
  \item \verb|harness/api/DisentangledOutlineJudgeBlueprintAPIService.py|
  \item \verb|harness/api/SummaryQAGeneratorAPIService.py|
\end{itemize}

\textbf{Configuration:}
\begin{itemize}
  \item \verb|harness/config/test_harness_outline_xhs.gin|
\end{itemize}

\textbf{Note:} Evaluation service files
(\verb|JudgeSearchQueryDiversityAPIService.py|,
\verb|JudgeSearchQueryDiversity_ZH.jinja2|) are not in the snapshot
list because they must not be modified.

\textbf{Non-modifiable gin parameters:} The following two parameters
are fixed and must not be modified:
\begin{itemize}
  \item \verb|DisentangledOutlineJudgeBlueprintAPI.search_engine = "xiaohongshu"|
  \item \verb|DisentangledOutlineJudgeBlueprintAPI.num_searches = 10|
\end{itemize}

\#\#\# Saving a Snapshot (execute after each eval round)

\begin{enumerate}
  \item Create \verb|harness/optimization_runs/v{N}/snapshot/| directory
  \item Copy all files from the snapshot list into \verb|snapshot/|,
        preserving the relative path structure (e.g.,
        \verb|snapshot/template/xxx.jinja2|,
        \verb|snapshot/harness/api/xxx.py|)
  \item Locate the eval output directory and copy results:
  \begin{itemize}
    \item Eval output path: \verb|./outs/| (under project root)
    \item Directory name format:
          \verb|TEST_HARNESS_ZH_V2_batch_<YYYYMMDD_HHMMSS>/|
    \item Locate method:
          \verb|ls -td ./outs/TEST_HARNESS_ZH_V2_batch_* | head -1|
    \item Copy \verb|metrics.json|, \verb|details.jsonl|, and
          \verb|summary.md| from that directory to
          \verb|harness/optimization_runs/v{N}/|
  \end{itemize}
  \item Write or update \verb|changelog.md| (fill in actual results)
  \item Update \verb|manifest.json|
\end{enumerate}

\#\#\# Restoring a Snapshot (execute at the start of each round)

Copy all files from \verb|best_version/snapshot/| back to the
corresponding locations in the working directory. For example:

\begin{verbatim}
cp harness/optimization_runs/v1/snapshot/template/*.jinja2 template/
cp harness/optimization_runs/v1/snapshot/harness/api/*.py harness/api/
cp harness/optimization_runs/v1/snapshot/harness/config/*.gin \
   harness/config/
# After restoring, confirm that search_engine="xiaohongshu"
# and num_searches=10 have not been altered
\end{verbatim}

\noindent\rule{\linewidth}{0.4pt}

\#\# V. Changelog Format

Each optimization round \textbf{must write changelog.md before executing
changes}. Format:

\begin{verbatim}
# V{N} Optimization Notes

## Problems Identified
(Analyze the previous round's eval data; describe the search_coverage
issues with data and reasoning support)
- search_coverage mean value; which cases scored lowest
- Common patterns in search_coverage_reasoning for low-score cases
- Any anomalies in process metrics (query count, doc_avg_relevance, etc.)

## Changes Made

### Modified files
- `path/to/file`: what was specifically changed

## Expected Outcome
(Expected changes to search_coverage)

## Actual Outcome
(Fill in after eval completes)
- search_coverage: {before} -> {after}
- overall: {before} -> {after}
- completeness: {before} -> {after}
- diversity: {before} -> {after}
\end{verbatim}

\noindent\rule{\linewidth}{0.4pt}

\#\# VI. Optimization Space

The agent may freely decide the direction and extent of optimization.
Modifiable files are listed in the snapshot list in Section IV.

\textbf{The sole non-modifiable items are the evaluation services:}
\verb|JudgeSearchQueryDiversityAPIService.py| and
\verb|JudgeSearchQueryDiversity_ZH.jinja2|.

\#\#\# Diagnosis-to-Action Mapping

Based on the data and reasoning in \verb|summary.md|, once the problem
is located, take action according to the following mapping:

\medskip
\begin{tabular}{|p{3.0cm}|p{3.5cm}|p{5.2cm}|}
\hline
\textbf{Symptom} & \textbf{Possible Cause} & \textbf{Priority Target} \\
\hline
Many queries return low-relevance docs (low high\_relevance\_ratio)
  & Search queries too broad / abstract, weak connection to query
  & Search query generation guidelines in outline template;
    \verb|_generate_additional_queries| logic in API \\
\hline
All search queries under a blueprint are low quality
  & Blueprint decomposition drifts from core query intent
  & Blueprint update strategy in outline template;
    \verb|IntentPlanner_ZH.jinja2| \\
\hline
Insufficient search query count (low search\_query\_count)
  & \verb|min_query_per_blueprint| or \verb|min_query_len| too low
  & Gin configuration parameters \\
\hline
Many duplicate / near-synonym queries
  & Insufficient deduplication logic; prompt does not emphasize diversity
  & \verb|_extract_search_queries| in API; search query guidelines in
    outline template \\
\hline
Low doc\_count (few documents returned)
  & Search queries too niche / specialized / long
  & Outline template (add broad+narrow mix);
    \verb|_broaden_queries|, \verb|_truncate_long_queries| in API \\
\hline
completeness\_reasoning points to missing dimensions
  & Incomplete blueprint coverage
  & Outline template (diversity requirements section);
    \verb|IntentPlanner_ZH.jinja2| \\
\hline
doc\_avg\_relevance high but search\_coverage low
  & A few queries severely drag down the overall score; or snippets
    are semantically highly overlapping
  & Identify low-quality queries; if snippets overlap, increase
    differentiation between search queries \\
\hline
doc\_avg\_relevance generally high ($>$0.6) but snippets do not
actually support the query
  & SummaryQA scoring too lenient: docs related to blueprint but not
    to user question received high scores
  & \verb|template/SummaryQAGeneratorBlueprint_ZH.jinja2| ---
    tighten judge scoring criteria; emphasize direct relevance to
    \textbf{user question} \\
\hline
doc\_avg\_relevance generally low ($<$0.3)
  & SummaryQA scoring too strict; or search results are genuinely
    not relevant
  & First check search query quality; if queries are reasonable but
    scores are low, read \verb|SummaryQAGeneratorBlueprint_ZH.jinja2|
    to check if scoring criteria are too strict \\
\hline
Supplementary queries generated by \verb|_generate_additional_queries|
are low quality
  & Regex-based supplementation strategy generates irrelevant queries
  & Rewrite \verb|_generate_additional_queries| or adjust
    \verb|min_query_per_blueprint| to avoid triggering it \\
\hline
\end{tabular}

\medskip
\#\#\# Strategy 1: Search Query Quality Optimization
\textit{(most direct impact on search\_coverage)}

Search queries are the \textbf{direct driver} of
\verb|search_coverage|. Checklist:
\begin{itemize}
  \item \textbf{Specificity:} Does each search query have a clear
        information-retrieval target? Avoid broad queries like
        ``how is XX''
  \item \textbf{Relevance:} Are the documents returned by the search
        query relevant to the \textbf{user query}? (Being relevant to
        a blueprint alone is not enough; it must relate to the original
        query)
  \item \textbf{Searchability:} On the target search engine
        (Xiaohongshu), can this query retrieve effective content?
\end{itemize}

\textbf{Entry points:}
\begin{itemize}
  \item \verb|template/DisentangledOutlineJudgeBlueprintStyleQA_ZH.jinja2|
        --- search query generation guidelines (most commonly modified)
  \item \verb|harness/api/DisentangledOutlineJudgeBlueprintAPIService.py|
        --- \verb|_validate_and_fix_blueprints|,
        \verb|_generate_additional_queries|,
        \verb|_check_query_quality|
\end{itemize}

\#\#\# Strategy 2: Search Query Count and Coverage

Does the total number of search queries cover all dimensions of the
query? Is the number of queries per blueprint sufficient?

\textbf{Entry points:} gin configuration (\verb|min_query_len|,
\verb|min_query_per_blueprint|, \verb|max_query_len|)

\#\#\# Strategy 3: Outline Structure Optimization

Blueprints determine the direction of search queries. If blueprints
deviate from the core intent of the query, even high-quality search
queries will be futile. When \verb|completeness_reasoning| repeatedly
points to missing dimensions, prioritize fixing this.

\textbf{Entry points:}
\begin{itemize}
  \item \verb|template/DisentangledOutlineJudgeBlueprintStyleQA_ZH.jinja2|
        --- outline key point update strategy
  \item \verb|template/IntentPlanner_ZH.jinja2|
        --- intent analysis prompt
\end{itemize}

\#\#\# Strategy 4: Search Parameter Tuning

Adjustable parameters such as \verb|top_k|, filter strategies, etc.
(Note: \verb|search_engine| and \verb|num_searches| cannot be modified.)

\textbf{Entry points:} gin configuration

\#\#\# Strategy 5: Post-Processing and Quality Control

Deduplication, truncation, expansion, and quality-checking logic applied
after search query generation. When analysis reveals that low-quality
queries originate from post-processing (e.g., queries supplemented by
\verb|_generate_additional_queries|, queries expanded by
\verb|_broaden_queries|), prioritize fixing here.

\textbf{Entry points:}
\begin{itemize}
  \item \verb|harness/api/DisentangledOutlineJudgeBlueprintAPIService.py|
        --- \verb|_smart_split_query|, \verb|_truncate_long_queries|,
        \verb|_broaden_queries|, \verb|_check_query_quality|
\end{itemize}

\#\#\# Strategy 6: Search Query Effectiveness Feedback Mechanism

The Memory system extracts the actual retrieval effectiveness of search
queries for the same query from historical traces, and injects this
information into the pipeline LLM's prompt at runtime to help it generate
more effective search queries.

\textbf{Design principles:}
\begin{itemize}
  \item \textbf{Exact query matching only:} Optimal search strategies
        vary greatly across different query types; fuzzy / global
        matching introduces noise
  \item \textbf{Feed back search query effectiveness data only, do not
        expose old-round blueprints:} Avoids anchoring to old-round
        outline structures; template instructions (not historical
        examples) should drive blueprint generation
  \item \textbf{Aggregate cross-round data:} The effectiveness of all
        search queries for the same query across multiple rounds is
        aggregated to form a whitelist/blacklist of search queries for
        that query
\end{itemize}

\textbf{Working mechanism:}
\begin{enumerate}
  \item After each batch eval round, automatically save
        \verb|harness/memory/traces_{timestamp}.jsonl|, containing
        per-query search query document quality details
        (\verb|per_query_doc_stats|: \verb|avg_relevance|,
        \verb|high_relevance_ratio|, etc. per query)
  \item At the start of the next round, for each query, exactly match
        historical traces and aggregate effectiveness data across all
        historical search queries (when the same query appears in
        multiple rounds, use the most recent data)
  \item Divide into high-effectiveness / low-effectiveness groups
        using a \verb|high_relevance_ratio| threshold of 0.5 (aligned
        with the judge scoring criterion ``relevance score $>$ 0.5'')
  \item Format and inject into the user prompt at \verb|turn_id=0|.
        The system prompt of the outline generation template contains
        ``Historical Search Query Effectiveness Feedback Usage
        Instructions'' to guide the LLM in utilizing the data
\end{enumerate}

\textbf{Example injected content:}

\begin{verbatim}
## High-effectiveness queries (3, returned documents highly relevant)
- "XX brand review": avg_relevance=0.72, 8/10 documents relevant

## Low-effectiveness queries (2, please avoid similar queries)
- "how is XX": avg_relevance=0.15, 1/10 documents relevant
\end{verbatim}

\textbf{Fallback:} If historical traces have no per-query details
(old format), only show the search queries and evaluations from
low-scoring rounds as negative references when the score gap $\geq$ 1.0.

\textbf{Adjustable directions:}
\begin{itemize}
  \item Display count limit (currently 5 per group; modify
        \verb|_format_query_performance_feedback|)
  \item High / low effectiveness threshold (currently 0.5; modify
        \verb|_format_query_performance_feedback|)
  \item Fallback score gap threshold (currently 1.0; modify
        \verb|_format_overall_feedback|)
\end{itemize}

\textbf{Note:} During the first batch eval (v0\_baseline), the memory
directory is empty and no feedback will be injected. Takes effect from
v1 onward. Since \verb|fixed_sample_size=10| guarantees the same set
of queries is used across rounds, exact matching is guaranteed to hit.

\medskip
\#\#\# Data-Driven Analysis

\begin{itemize}
  \item \textbf{summary.md} (must read each round): aggregate metrics,
        statistics by intent, full case details (including blueprints,
        search queries, \textbf{all-dimension reasoning},
        \textbf{per-query search query quality distribution and
        low-quality query details})
  \item \textbf{details.jsonl}: per-case metric details
  \item \textbf{harness/memory/traces\_*.jsonl}: complete trace for each
        round (including per-query document quality details), for
        deeper analysis
  \item \textbf{full\_results.jsonl} (in the batch output directory):
        complete \verb|run_agent| output with all intermediate results
\end{itemize}

Do not limit yourself to minor parameter adjustments. If analysis
reveals a structural problem, the modification should be of corresponding
magnitude.

\noindent\rule{\linewidth}{0.4pt}

\#\# VII. Precautions

\begin{enumerate}
  \item \textbf{Working directory:} All commands are executed under the
        project root directory.
  \item \textbf{Do not modify} \verb|run_batch_agent|,
        \verb|extract_metrics|, or \verb|aggregate_metrics|
        \textbf{in} \verb|eval_outline_judge.py|.
        Only modify templates, API service code, and configuration.
  \item \textbf{Evaluator not modifiable:}
        \verb|JudgeSearchQueryDiversityAPIService.py| and
        \verb|JudgeSearchQueryDiversity_ZH.jinja2| must not be
        modified. These are the final scoring services --- modifying
        the scorer is equivalent to cheating.
  \item \textbf{Each round's modifications must be based on
        best\_version only:} Do not continue modifying on top of a
        regressed version.
  \item \textbf{Read historical changelogs to avoid repetition:} Do not
        retry the same changes that have already failed.
  \item \textbf{Read summary.md before analyzing:} Before each round's
        analysis, first read \verb|summary.md| under the
        \verb|best_version| directory to understand the actual model
        output (blueprints, search queries, reasoning); do not rely
        solely on aggregate scores.
  \item \textbf{Read the code before modifying:} Before modifying any
        file, read the complete current version of the code to
        understand the existing logic.
  \item \textbf{API cost awareness:} Each batch round involves calls to
        Gemini + Google Search + GPT-oss.
  \item \textbf{Template paths:} All templates are loaded at runtime
        from \verb|./template/|. When modifying templates, edit files
        under \verb|template/|; snapshots also store files from
        \verb|template/|.
  \item \textbf{Parameter configuration:} All API service parameters in
        \verb|run_agent| are controlled via gin configuration
        (\verb|test_harness_outline_xhs.gin|). To adjust parameters,
        modify the gin configuration file.
  \item \textbf{Non-modifiable search parameters:}
        \verb|search_engine = "xiaohongshu"| and
        \verb|num_searches = 10| are fixed and must not be modified.
  \item \textbf{Optimization target:} Only \verb|search_coverage.mean|
        is the optimization target; all other metrics are for reference
        and display purposes only.
\end{enumerate}

\end{tcolorbox}

\subsection{Prompts in Harness}
\label{app:harness_prompt}
As introduced in Section~\ref{sec:arch}, the outline generator agent is repurposed as a scorer agent, the prompt of which is presented as follows.

\begin{tcolorbox}[promptbox, title=\small Prompt: Scorer Agent]
\small

\# Role Definition

You are a professional expert in information retrieval and content
planning evaluation, specializing in assessing the completeness and
diversity of blueprints and search-queries in covering user queries,
and capable of performing quality verification based on real retrieval
result distribution data.

\noindent\rule{\linewidth}{0.4pt}

\#\# Input Description

\begin{itemize}
  \item \textbf{User query:} the user's original needs and questions
  \item \textbf{blueprints:} a list of outline points, each containing:
  \begin{itemize}
    \item \textbf{content:} description of the point
    \item \textbf{search\_query:} the corresponding list of search
          queries
  \end{itemize}
  \item \textbf{Search result distribution data:} primarily the
        relevance judgment scores between documents returned for the
        search queries and the user query
\end{itemize}

\noindent\rule{\linewidth}{0.4pt}

\#\# Evaluation Dimensions

\#\#\# Evaluation Dimension 1: Completeness of blueprints and
search-queries (completeness)

\textbf{Scoring perspectives:}
\begin{enumerate}
  \item \textbf{Topic coverage:} Do the blueprints and search-queries
        cover all core topics and sub-topics required to answer the
        query?
  \item \textbf{Intent recognition:} Do they capture both the user's
        explicit intent (what is directly asked) and implicit intent
        (underlying purpose)?
  \item \textbf{Dimension diversity:} Are multiple dimensions explored
        (e.g., background, current status, causes, impact, comparison,
        trends, recommendations, etc.)?
  \item \textbf{Missing key points:} Identify important content points
        involved in the query but not covered by the blueprints and
        search-queries.
\end{enumerate}

\textbf{Scoring criteria (0--10):}
\begin{enumerate}
  \item 9--10: Fully covers all core dimensions of the query, no
        obvious omissions, well-balanced dimension distribution
  \item 7--8: Covers most core dimensions, with 1--2 minor omissions
  \item 5--6: Covers basic dimensions, but with more than 1 important
        dimension missing
  \item 3--4: Limited coverage, multiple important dimensions missing
  \item 1--2: Covers only a very small amount of relevant content,
        severely insufficient
  \item 0: Completely fails to cover the content required by the query
\end{enumerate}

\#\#\# Evaluation Dimension 2: Diversity of blueprints and
search-queries (diversity)

\textbf{Scoring perspectives:}
\begin{enumerate}
  \item \textbf{Perspective diversity:} Do the blueprints and
        search-queries explore different standpoints / roles / group
        perspectives (user perspective, expert perspective, policy
        perspective, market perspective, etc.)?
  \item \textbf{Content type diversity:} Do they simultaneously cover
        factual content, analytical content, comparative content, and
        advisory content?
  \item \textbf{Granularity diversity:} Is there both a macro framework
        and micro-level details?
  \item \textbf{Redundancy:} Is there a large amount of overlapping or
        semantically similar content among the blueprints and
        search-queries?
\end{enumerate}

\textbf{Scoring criteria (0--10):}
\begin{enumerate}
  \item 9--10: Rich and diverse perspectives, comprehensive content
        types, reasonable granularity distribution, no obvious
        repetition
  \item 7--8: Good diversity, with occasional perspective repetition or
        type singularity
  \item 5--6: Average diversity, with one type or perspective being
        overly dominant
  \item 3--4: Insufficient diversity, content is highly homogeneous
  \item 1--2: Severely lacking in diversity, large amount of repetition
  \item 0: Completely no diversity
\end{enumerate}

\#\#\# Evaluation Dimension 3: Search-query retrieval quality
(search coverage)

\textbf{Scoring perspectives:}
\begin{enumerate}
  \item \textbf{Retrieval effectiveness verification:} Verified based
        on the distribution of relevance judgment scores of retrieved
        content
  \item \textbf{Cross-query result overlap:} Whether the semantics of
        snippets from different retrieved contents are complementary
\end{enumerate}

\textbf{Scoring criteria (0--10):}
\begin{enumerate}
  \item 9--10: All search-query information is sufficiently supported;
        the majority (above 80\%) of search results are highly relevant
        (relevance score $>$ 0.5)
  \item 7--8: The majority (above 80\%) of search-query content is
        sufficient, with the information chain basically complete
  \item 5--6: 20\%--40\% of search-query content is insufficient, with
        information gaps present
  \item 3--4: More than 40\% of search-query content is insufficient or
        directionally off
  \item 1--2: Most retrieval results are unable to support the
        corresponding search-queries
  \item 0: Completely no relevant content
\end{enumerate}

\noindent\rule{\linewidth}{0.4pt}

\#\# Output Format

Please strictly output the evaluation results in the following JSON
format:

\begin{verbatim}
{
  "evaluation": {
    "completeness": {
      "score": 0,
      "reasoning": "explanation of scoring rationale"
    },
    "diversity": {
      "score": 0,
      "reasoning": "explanation of scoring rationale"
    },
    "search_coverage": {
      "score": 0,
      "reasoning": "explanation of scoring rationale"
    },
    "overall": {
      "score": 0,
      "reasoning": "overall evaluation summary"
    }
  }
}
\end{verbatim}

\end{tcolorbox}

\begin{figure}
    \centering
    \includegraphics[width=1\linewidth]{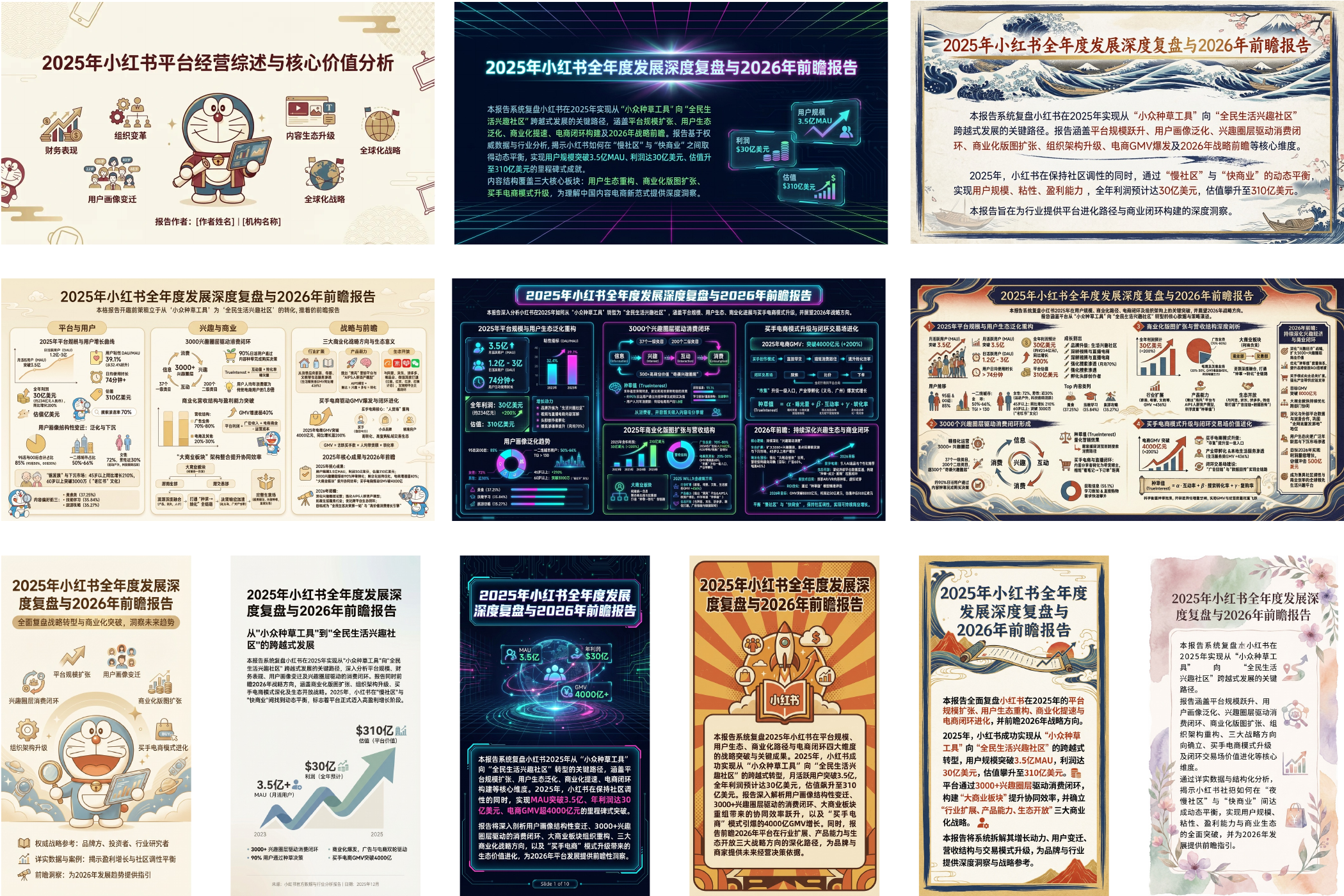}
    \caption{\textbf{Gallery of diverse template styles and types}, including slides, posters, and portrait-format images. Our render agent offers extensive stylistic choices, accommodating diverse user preferences. }
    \label{fig:agentdisco_templates}
\end{figure}

\section{Gallery of Templates}
\subsection{Gallery of Templates in Render Agent}
\label{app:showcase_render}
As described in Section~\ref{sec:render}, the render agent offers a diverse collection of plug-in style templates, representative examples of which are illustrated in Figure~\ref{fig:agentdisco_templates}.

\end{document}